\documentclass[]{aa} 

\usepackage{graphicx}
\usepackage{txfonts}
\usepackage[dvipsnames]{xcolor}
\usepackage{natbib}
\usepackage{longtable}
\usepackage{siunitx}
\usepackage{enumerate}
\usepackage{lscape}
\usepackage{hyperref}
\usepackage[normalem]{ulem}

\hypersetup{
  colorlinks=true,   
  urlcolor=blue,     
  linkcolor=blue, 
  citecolor = blue
}

\makeatletter
\renewcommand*\aa@pageof{, page \thepage{} of \pageref*{LastPage}}
\makeatother

\bibpunct{(}{)}{;}{a}{}{,}    

\newcommand{\lint}{$L_{\rm int}$\,} 
\newcommand{\lbol}{$L_{\rm bol}$\,} 
\newcommand{\tbol}{$T_{\rm bol}$\,} 
\newcommand{\lo} {$L_{\odot}$\,}
\newcommand{\mj} {$M_\mathrm{Jup}$}
\newcommand{\mo} {$M_{\odot}$}
\newcommand{\Menv}{$M_\mathrm{env}$}

\begin{document}

\title{Substellar candidates at the earliest stages: the SUCANES database}

\author{
A.M. P\'erez-Garc\'{\i}a\inst{1}
\and
N. Hu\'elamo\inst{2}
\and
A. Garc\'{\i}a L\'opez\inst{1}
\and
R. P\'erez-Mart\'{\i}nez\inst{1}
\and
E. Verdugo\inst{3}
\and
Aina Palau\inst{4}
\and
I. de Gregorio-Monsalvo\inst{5}
\and
O. Morata\inst{6}
\and
D. Barrado\inst{2}
\and
M. Morales-Calderon\inst{2}
\and
M. Mas-Hesse\inst{2}
\and
A. Bayo\inst{7}
\and
K. Mauc\'o\inst{7}
\and
H. Bouy\inst{8}
}
\institute{
ISDEFE Beatriz de Bobadilla 3, 28040 Madrid, Spain\\
 \email{amperez@isdefe.es}
 \and
Centro de Astrobiolog\'{\i}a (CAB), CSIC-INTA, ESAC Campus, Camino bajo del Castillo s/n, E-28692 Villanueva de la Ca\~nada, Madrid, Spain           
\and
European Space Agency, ESAC, Camino Bajo del Castillo, E-28692, Villanueva de la Ca\~nada, Madrid, Spain
\and
Universidad Nacional Aut\'onoma de M\'exico, Instituto de Radioastronom\'ia y Astrof\'isica, Antigua Carretera a P\'atzcuaro 8701, Ex-Hda. San Jos\'e de la Huerta, 58089 Morelia, Michoac\'an, M\'exico
\and
European Southern Observatory, Alonso de Cordova 3107, Casilla 19, Vitacura, Santiago, Chile
\and 
Institut de Ci\`encies de l'Espai (ICE-CSIC), Campus UAB, Can Magrans S/N, E-08193 Cerdanyola del Vall\`es, Spain
\and
European Southern Observatory, Karl-Schwarzschild-Strasse 2, 85748 Garching bei M\"unchen, Germany
\and
Laboratoire d’Astrophysique de Bordeaux, Univ. Bordeaux, CNRS, B18N, allée Geoffroy Saint-Hilaire, 33615 Pessac, France
}

\date{Received ---; accepted ---}
    

    \abstract
    {Brown dwarfs are the bridge between low-mass stars and giant planets. 
    One way of shedding light on their dominant formation mechanism is to study them at the earliest stages of their evolution, when they are deeply embedded in their parental clouds. Several works have identified pre- and proto-brown dwarfs candidates using different observational approaches.
    }
    {The aim of this work is to create a database with all the objects classified as very young substellar candidates in the litearature in order to study them in an homogeneous way.}
    {We have gathered all the information about very young substellar candidates available in the literature until 2020. We have retrieved their published photometry from the optical to the centimeter regime, and we have written our own codes to derive their bolometric temperatures and luminosities, and their internal luminosities. We have also populated the database with other parameters extracted from the literature, like e.g. the envelope masses, their detection in some molecular species, and presence of outflows.}
    {The result of our search is the SUCANES database, containing 174 objects classified as potential very young substellar candidates in the literature. We present an analysis of the main properties of the retrieved objects. Since we have updated the distances to several star forming regions, this has allowed us to reject some candidates based on their internal luminosities. We have also discussed the derived physical parameters and envelope masses for the best substellar candidates isolated in SUCANES.  
    As an example of a scientific exploitation of this database, we present a feasibility study for the detection of radiojets with upcoming facilities: the ngVLA and the SKA interferometers. 
    The SUCANES database is accessible through a Graphical User Interface and it is open to any potential user.}
    {}


   \maketitle

\nolinenumbers
\section{Introduction} \label{sec:1}

Brown dwarfs (BDs) are substellar objects with masses between $\sim$80-13\,$M_{\rm Jup}$, so they are the bridge between low-mass stars and Jupiter-like planets. 
A large number of observational works \citep[e.g.][]{Natta2002,Whelan2005,Barrado2007,Alves_Ol_2010,Bayo2012,Muzic2014} have shown that young BDs ($\sim$ 1-10 Myr) share properties with young low-mass stars, like e.g. the presence of disks and jets, suggesting that they share the same formation mechanism. 
However, since their discovery, the formation of BDs have been under debate,  aiming at understanding the dominant mechanism of substellar formation. The most widely discussed mechanisms are turbulent fragmentation \citep{padoan04,hennebelle08}, the ejection from multiple protostellar systems \citep{Reipurth01,Bate12}, highly erosive outflows \citep{Machida09} and disk fragmentation \citep{Whitworth06}. In the surroundings of high-mass stars, there are additional mechanisms that might form BDs, namely photo-evaporation of cores near massive stars \citep{Whitworth04}, and gravitational fragmentation of dense filaments formed in a nascent cluster \citep{Bonnell08}.
                                    
Since stars and BDs evolve very rapidly during the first million years, one way to shed light on the dominant BD formation scenario is to study the properties of BDs at the earliest stages of their evolution, when they are still embedded in the parental clouds. 
If BDs are a scaled-down version of low-mass stars, they are expected to form in isolation within molecular clouds with the same observational properties and correlations in physical parameters as those found in protostars (e.g. envelopes, molecular outflows, or thermal radiojets). By analogy with low-mass stars, BDs would be expected to show different evolutionary stages characterized by different physical and morphological properties, and spectral energy distributions \citep[SEDs; see][]{Lada1984, Lada1987, Andre1993_Class0}:  in the earliest phases we would find the pre-BDs, which are dense cores that will form a BD in a future but have not formed an hydrostatic core yet (i.e., it is an analog to the pre-stellar cores for low-mass stars). Class~0 proto-BDs would consist on embedded cold cores with SEDs peaking at the (sub-)mm regime, while Class~I proto-BDs should be characterized by an infalling envelope feeding an accreting circumstellar disk, and with SEDs  peaking in the far-infrared. 

From the observational point of view, several efforts have been made to identify the youngest BDs.
In the early 2000, the {\em Spitzer} Space Telescope identified a new family of objects called very low-luminosity objects \citep[VeLLOs][]{Young04,diFrancesco07}, that  were  recognized as potential  proto-BDs. They are deeply embedded objects characterized by internal luminosities\footnote{The internal luminosity is defined as the total luminosity arising from a central object and its disk (including both accretion and photospheric luminosity) and excluding any luminosity arising from external heating of the surrounding dense core by the interstellar radiation field.}  ($L_{\rm int}$) below 0.1-0.2\,$L_{\odot}$ \citep{Dunham08,Kim2016}. 
A number of these VeLLOs have been further characterized \citep[e.g.,][]{Bourke06,Dunham08,Palau12,Palau14, Morata15, Kim2016,Kim2019} using different observations, and display properties consistent with proto-BDs.

                  
Another group of objects with very low internal luminosities are the so-called First Hydrostatic Cores (FHCs). They are
supposed to form during the first stages of collapse, once the density is large enough to turn the collapse from isothermal to adiabatic, providing the required pressure to balance gravity. The predicted properties of FHCs are low internal luminosities, very low accreted masses, SEDs peaking around 100\,$\mu$m and association with low-velocity outflows. Thus, the properties of FHCs are similar to the properties expected for deeply embedded Class~0-like proto-BD; it is the mass reservoir in the envelope that is the key distinguishing parameter between the two: for a FHC to end-up as a substellar object, the mass of the envelope should be small enough to guarantee that the accreted material does not place the object in the stellar regime.
Only a few of FHC candidates have been identified so far \citep[see e.g.][]{Palau14, Hirano2019_B1bNS, Maureira2020_FHCs}. 

Pre-brown dwarf (pre-BD) cores are also potential substellar objects: they are cores detected mainly at continuum (sub-)mm wavelengths with no infrared/optical counterparts, and with \lint upper limits $<$ 0.2\,\lo; they are gravitationally bound but have very low masses (near substellar) and thus, they will not be able to form a stellar object even if accreting all the reservoir of mass within the core. So far, there is only one well characterized pre-BD \citep{Andre12}, and several candidates \citep[e.g.][]{Palau12,deGregorio2016,Huelamo17,Santamaria21}. 


As it follows, there are several studies that have identified substellar candidates at very early stages of their formation (FHCs, pre- and proto-BD) in different star forming regions
using different observational approaches.
A comprehensive study comparing the properties of all the candidates in an homogeneous way is needed in order to shed  light on their formation mechanism.   In this context, and as a first step, we have created the SUbstellar CANdidates at the Earliest Stages (SUCANES) database (DB): it is intended to be a compilation of all the sources that have been identified as probable pre-BD, proto-BD, VeLLO and FHC candidates in the literature. The SUCANES DB is accessible through a Graphical User Interface (GUI) and is open to any potential user.

In this manuscript, we describe how we built SUCANES, the  data that have been used to populate it, and include a brief analysis of its contents. A description of the sample selection and the data used in the DB are described in Section~\ref{db1}, while the GUI to access the DB is described in Section~\ref{GraphInt}. A summary of the main properties of the compiled objects is presented in Section~\ref{Contents}. Finally, we have included a scientific application of the DB in Section~\ref{sec:SciApp}. Our main conclusions are summarized in the last section of this paper. 
               
\section{The SUCANES database}\label{db1}

SUCANES has been built using data from the literature published until 2020 (included) and complemented with photometry from public archives and catalogues (from VizieR). 
The objects that populate SUCANES are those with these properties:
      \begin{itemize}
            \item Published works have classified the objects as pre- or proto-BDs
            \item Published works have classified the objects as VeLLOs, that is, young objects with low internal luminosities (\lint\,$<$ 0.2\,$L_{\odot}$, within the errors).
            \item Objects classified as FHCs.
            \item All the objects selected above show Class~0, 0/I, I, Flat, or an unknown class, that is,  we have not included Class II objects.

      \end{itemize}

The SUCANES database is constructed in a MariaDB\footnote{https://mariadb.org/en/} system, using python scripts. The database has been built creating different tables (in csv format) including different information about the targets.  All these tables are connected by a common object identification number (ObjID). 

The DB contains a total of eleven tables, that can be grouped in four different blocks: identification tables (2), photometric tables (6), physical parameters tables (2), and molecular line emission table (1). We describe all these tables and the information contained in each of them in the following subsections and in Appendix~\ref{ap_tab}.  

Note that all the SUCANES material is available in the
GitHub repository.

\subsection{SUCANES identification tables}\label{sec:iden_tables}


The first two tables of SUCANES are the identification tables, named Identity and Position, which contain basic information about the objects:

- {\bf Identity} includes two columns with the \texttt{Name} of the object as found in the literature, and \texttt{Other Name}, which is a resolver ID that facilitates archival searches. For this second column, we have used sesame resolvable IDs whenever available\footnote{http://vizier.cds.unistra.fr/vizier/doc/sesame.htx}, and the ViZieR association when the former did not exist.

- {\bf Position} includes the object coordinates, \texttt{RA(J2000)} and \texttt{DEC(J2000)}, the star forming region they belong to (\texttt{Region}), their \texttt{distance} (in pc), their  \texttt{Class} and \texttt{Type},  and the corresponding bibliographic references from which all the information, including the \texttt{Name} from the Identity table,  has been extracted. 

\begin{table}
\caption{Adopted distances for different regions in SUCANES}
    \centering
    \begin{tabular}{lrl}\hline
     Region    & Distance & Reference\\
               & [pc]    \\ \hline
    Aquila & 436$\pm$9 & \citet{OrtizLeon2018a}\\
    Cepheus/L1251      &  339$\pm$1 & \citet{Szil2021}\\
    Cepheus/L1148      &  330$\pm$1 & \citet{Szil2021}\\
    Cepheus/NGC7023    &  341$\pm$2 & \citet{Szil2021}\\
    Cha I      &  192$\pm$6 & \citet{Dzib2018}\\
    Cha II     &  198$\pm$6& \citet{Dzib2018}\\
    Cha III     & 193$\pm$12 & \citet{Voirin2018}\\
    Corona Australis & 154$\pm$4 & \citet{Dzib2018}\\
    GF9 & 270$\pm$10 & \citet{Clemens2018}\\
    IC\,348          & 321$\pm$10 & \citet{OrtizLeon2018b}\\
    IC\,5146    & 600$\pm$100 & \citet{Wang2020} \\
    Ophiuchus/L1688 & 138.4$\pm$2.6 & \citet{OrtizLeon2018a}\\
    Ophiuchus/L1689 & 144.2$\pm$1.3  & \citet{OrtizLeon2018a}\\
    Perseus         &  293$\pm$22&  \citet{OrtizLeon2018b}\\
    Serpens         & 436$\pm$9 & \citet{OrtizLeon2018a}\\
    Sigma Ori & 402$\pm$25 & \citet{Monteiro2020}\\ 
    \hline
    \end{tabular}
    \label{tab:distances}
\end{table}

 We have considered updated distances to different star forming regions following the results from works using mainly Gaia and Very Long Baseline Array (VLBA) data (see Table~\ref{tab:distances} for the correspoding references): we have adopted new distance values to Perseus, IC348, Serpens, Aquila, GF\,9, Corona Australis, and $\sigma$ Orionis. In the case of Ophiuchus objects,  we have adopted different values for objects in the  L\,1689 and L\,1688 clouds. 
 IC\,5146 has been studied by \citet{Wang2020}, concluding that the region has two components: the Cocoon Nebula and the filaments. After checking that all the objects in SUCANES lie in the filaments, we have adopted a distance of 600$\pm$100\,pc. In the case of Cepheus, we have compared the location of the nine objects included in SUCANES within the groups defined by \citet{Szil2021}: three lie in  the NGC\,7023,  and four in L\,1251.
 One of the objects (J215607) lies close to this latter region, but outside the area defined in that work. Hence, we have included 'Cepheus/L1251?' in its 'Region' field, while the ninth object lies in the L1148 cloud.  Finally, for Chamaeleon we have adopted different distances for the I, II and III clouds.
For the rest of the regions, we have used the distances quoted in the different publications: some of them have been already estimated using Gaia data \citep[e.g. Lupus,][]{Santamaria21}, or are consistent with those reported in recent studies including Gaia data \citep[e.g. Taurus, Orion, California; see][]{Galli2019,Zucker2020}. 

The \texttt{Class} column refers to the evolutionary class of the objects \citep[see e.g.][]{Lada1984, Lada1987}. 
This class has been traditionally estimated measuring the slope ($\alpha$$_{\rm IR}$) of the SED in the infrared range (normally from 2 to 10-25\,$\mu$m), so that the objects follow an evolutionary sequence covering several classes: I, Flat, II, and III. An earlier class (Class~0) was defined later by \citet{Andre1993_Class0} based on different criteria.
Another approach to derive the evolutionary stage of a young stellar object is to use the bolometric temperature (\tbol) that comes from the SED (see section~\ref{subsec:phys}). The  connection between the \tbol value and the classes was first done by \citep{Chen1995}, and later revised by \citet{Evans2009_c2dLegacy}.
SUCANES includes objects with classes between 0 and Flat, 
that have been retrieved from the literature. Since we are studying embedded objects, the class has been always derived using the \tbol value. In the particular case of \citet{Kim2016}, this work estimated a \tbol value for the sources, but they did not explicitly mention a class. In those cases, we have assigned 
a value using the quoted \tbol and following the prescription by \citet{Chen1995}. 
Finally, note that the objects classified as pre-BD or FHCs candidates are objects less evolved than Class~0, so they are not within the classification  scheme described above. In these cases, we have filled the \texttt{Class} field with 'N.A.'.

The \texttt{Type} column refers to the classification of the objects in the literature, and its value can be 'proto-BD', 'pre-BD', 'FHC', or 'VeLLO'. In three cases, we have also included the type 'protostar' since the nature of the object is under discussion. 
It is important to remark that, regardless of this classification, all the objects in SUCANES should be considered as substellar candidates, since there are not unambiguous observations that have confirmed them as pre- or proto-BDs yet.
A brief discussion of individual objects whose nature is not clear is included in Appendix~\ref{ap_indiv_sources}.

Finally, note that we have included a 'D' after the type (e.g. 'VeLLO/D' and 'proto-BD/D')  to indicate objects discarded as young substellar candidates after our own analysis {\bf presented here}: they are mainly discarded after the recalculation of the internal luminosities using updated distances (see Sect. \ref{sec:lint}), or due to other findings described below, like e.g. new Euclid data indicating that the source is extragalactic.

\subsection{SUCANES photometric tables}

SUCANES contains photometric information (continuum data) of the objects from the optical to the centimeter range. We have created  a total of six tables with these information:

- {\bf Optical:} this table contains data in RIZ-filters. The table includes the measured magnitudes and the instrumentation used.

- {\bf NIR:} it contains photometry in the $JHKs$ filters. The table includes the measured magnitudes and the instrumentation used.

- {\bf Spitzer:} contains photometric data obtained with the IRAC \& MIPS instruments onboard the Spitzer space telescope. We note that we have only considered Spitzer and not WISE data for the wavelength range between 3-24 $\mu$m, given the better angular resolution of the former.

- {\bf Herschel:} contains photometric data obtained with the PACS \& SPIRE instruments onboard the Herschel telescope. 

- {\bf Sub-milimeter:} This table includes sub-mm continuum fluxes. It has been built using published data from single dish and interferometer facilities. It is the most inhomogeneous table, in the sense that contains data covering a broad wavelength range (from 350$\mu$m to 7.3\,mm), obtained at very different angular resolutions.  To provide as much as possible information, we have included in the table the observing wavelength, the flux density used to build the SED, the peak intensity and the integrated flux, the size and position angle (PA) of the beam, the angular size and PA of the source, and the name of the interferometer (or antenna) where the data was obtained. We have also included the flag 
\texttt{Resolved}, which 
adopts a value of '0' if the object is not spatially resolved, and a value of '1' if it is resolved. 


- {\bf Centimeter:} This table includes data in the centimeter regime, from 1\,cm longwards. The contents of the table are similar to the sub-millimeter one, and has been populated mainly with interferometric data from the Jansky Very Large Array (JVLA). 
In this table we have added an additional flag called \texttt{Variability}, which provides information of the source variability at 8.5\,GHz (3.5\,cm): we have populated this column with the results included in \citet{Choi2014_variability}, that studied the centimeter variability of ten VeLLOs. In case there is information about the source variability it adopts a value of '1', and if there is not information a value of '0'. 


The optical and near-IR photometry are provided in Vega magnitudes, while the rest of the fluxes are  provided in mJy. {\bf For clarity}, we have built an auxiliary table (named {\bf Filters}) with the filters used to obtain the photometric data, including the corresponding zero-points to change from magnitudes to fluxes, or viceversa\footnote{The Users Manual of the database also includes a link to the Spanish Virtual Observatory filter system for each filter in this table}. All the magnitudes/fluxes have an associated error. Whenever this error value is equal to 0.0, the corresponding magnitude/flux is an upper limit. Note that all the photometric tables include a column with the bibliographic references from which the data have been retrieved. We provide a detailed description of the contents of each photometric table in Appendix \ref{ap_tab}.

\begin{figure*}[t!]
    \centering
    \includegraphics[width=1\textwidth]{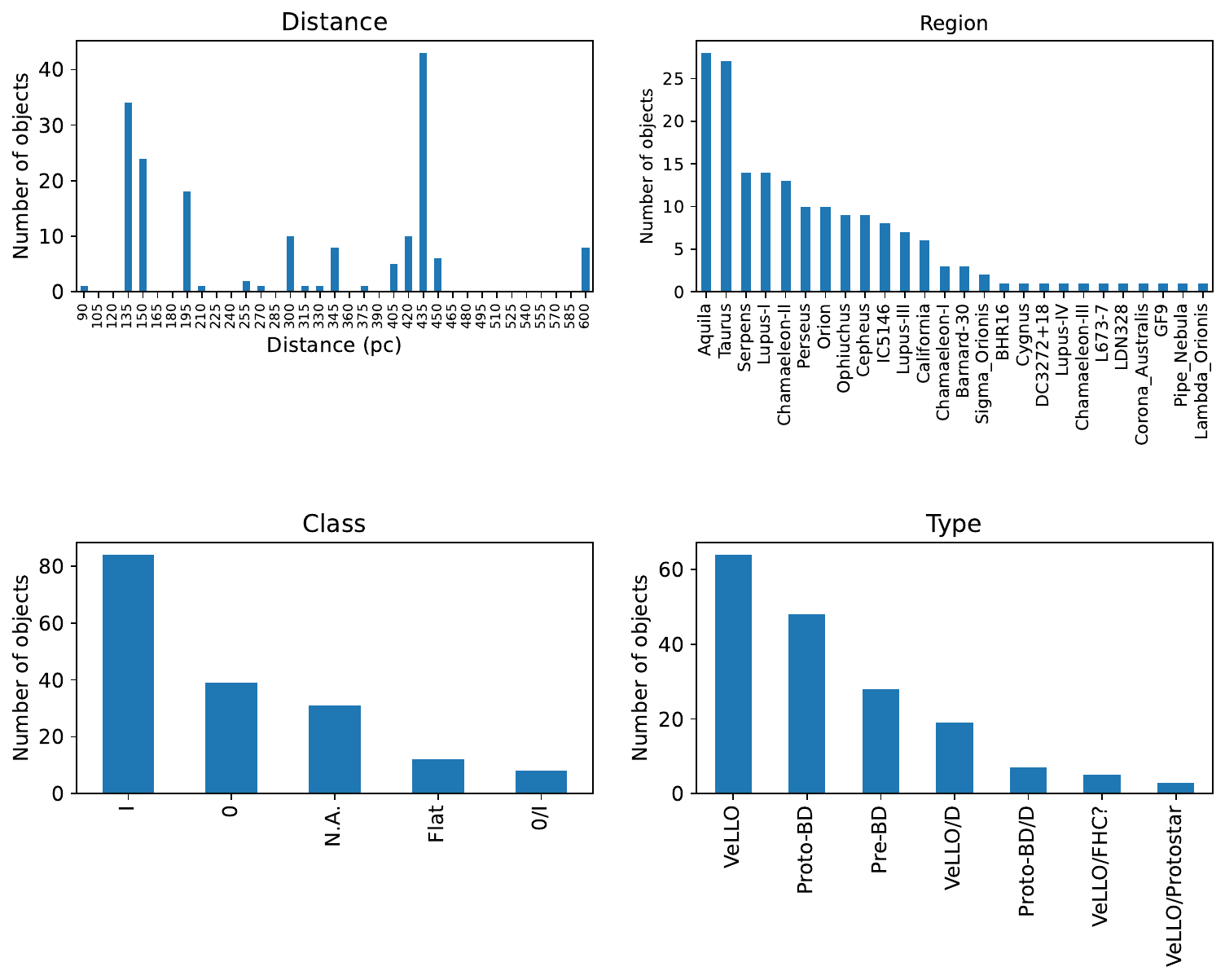}
 \caption{Histograms displaying the main properties of the objects included in SUCANES. The top panels show the distribution of distances and the star-forming region or associations the objects belong to. The bottom panels display the evolutionary stage of the objects through the derived 'Class' parameter (see Section~\ref{sec:iden_tables}), while the classification of the sources as derived in the different publications is shown in the 'Type' histogram.}
    \label{fig:hist1}
\end{figure*}
\begin{figure*}[t!]
    \centering
    \includegraphics[width=1\textwidth]{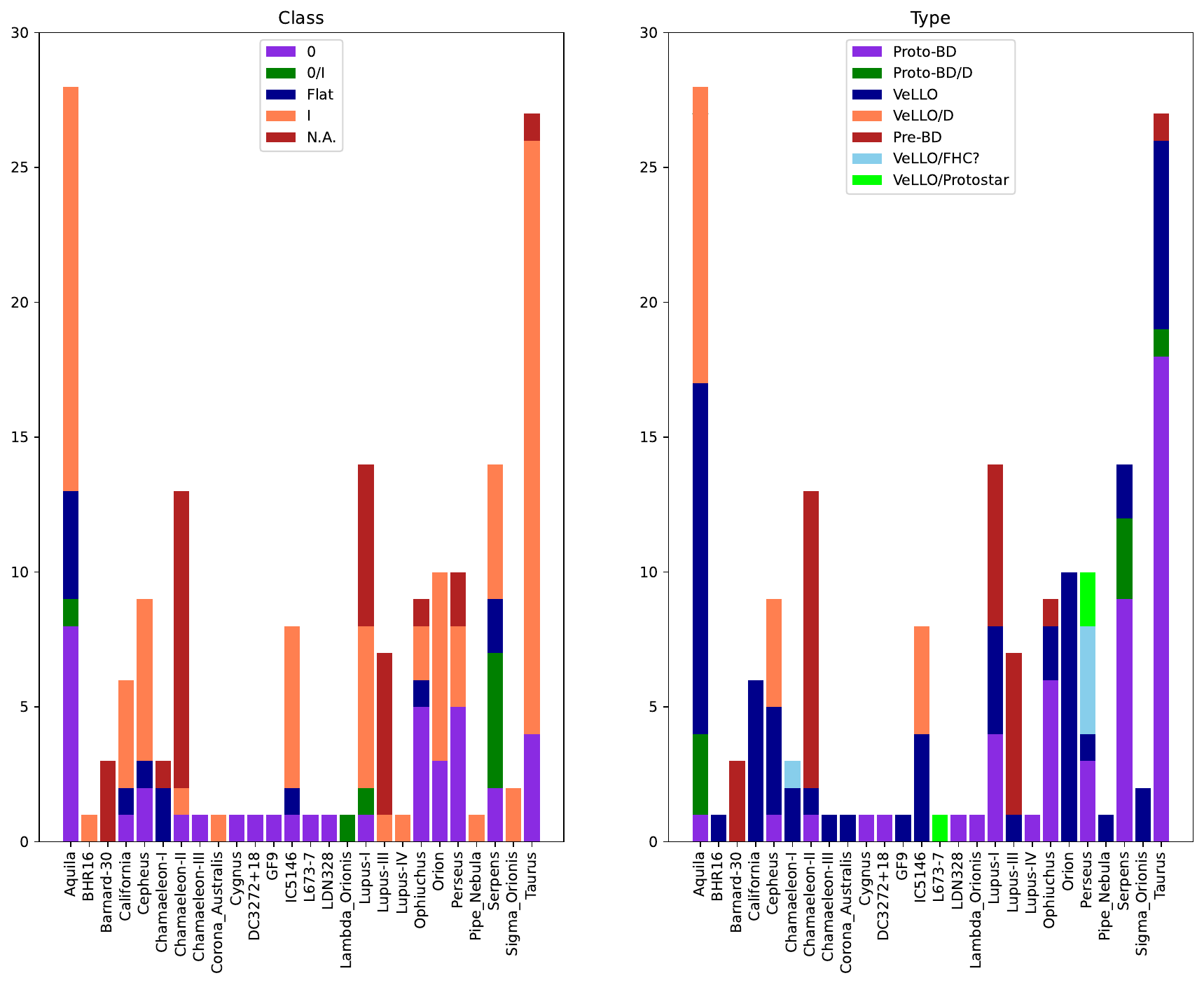}
\caption{Histogram of the SUCANES objects separated by star forming region. The left histogram displays the number of objects according to their 'Class', while the right histogram shows the distribution of objects according to their 'Type', as given in the literature.}
    \label{fig:hist2}
\end{figure*}

\subsection{SUCANES physical parameters tables}\label{subsec:phys}

The SUCANES DB includes two tables with physical parameters of the sources: {\bf LumTbol} and {\bf Dust\_properties}. 
The first table contains information about the internal luminosity ($L_{\rm int}$), bolometric luminosity ($L_{\rm bol}$), and bolometric temperature ($T_{\rm bol}$) of the objects, and the second table includes information about the dust mass ($M_{\rm dust}$) and total mass (gas+dust, $M_{\rm total}$) of the envelopes around the objects. While we have derived the parameters included in the first table using the retrieved photometry and our own codes implemented in the DB, the second table is populated with data from the literature. We describe their contents below.

{\bf LumTbol:} for this  table, we have estimated the physical parameters of the objects in the following way:

- {\em Internal Luminosity}: The internal luminosity is defined as the total luminosity arising from a central object and its disk (including both accretion and photospheric luminosity), and excluding any luminosity arising from external heating of the surrounding dense core by the interstellar radiation field. The internal luminosity (in units of solar luminosity) is estimated using the flux at  70\,$\mu$m (from PACS@Herschel or MIPS@Spitzer) and the distance to the source, {\it D} in pc, adapting the prescription by \citet{Dunham08}:

\begin{equation}\label{eq:lint}
L_{\mathrm{int}}\,(L_\odot)\, = \,3.3\times 10^8 \cdot [ F_{70} \times (D/140)^2]^{0.94}    
\end{equation}

where $F_{70}$ corresponds to $\lambda\cdot$$F_{\lambda}$ in erg/s/cm$^2$.


Given the better sensitivity and angular resolution of the Herschel/PACS data  in comparison with {\em Spitzer}/MIPS, we always consider the former  to derive \lint if both fluxes are available.

\citet{Dunham08} considered VeLLOs as those objects with $L_{\rm int} \leq$0.1\,$L_{\odot}$. However, other works have relaxed this limit to 0.2\,\lo\, \citep[see e.g.][]{Kim2016} to take into account a factor of $\sim$2 uncertainty associated to the $L_{\rm int}$ estimation from  the 70\,$\mu$m flux using Eq.~\ref{eq:lint}. We have adopted this latter approximation, and have considered VeLLOs those that fulfill that their \lint is $\leq$ 0.2\lo\, within the errors, that is, \lint\,- err\lint\,$\leq$ 0.2\,\lo. We discuss in detail the distribution of internal luminosities in Section~\ref{Contents}.
 

{\em - Bolometric luminosity}: The $L_{\rm bol}$ of any source in SUCANES has been estimated integrating its SED from the optical to the millimeter {\bf including detections and upper limits}, and then using its distance ({\it D}) to convert the flux into a luminosity. To calculate $L_{\rm bol}$, we have imposed that the object should be detected in, at least,  three different wavelength ranges (i.e., three different photometric tables) to ensure a significant spectral coverage. If the object does not fulfill this condition, the bolometric luminosity is not computed. 

{\em - Bolometric temperature:} The $T_{\rm bol}$ of an object with an spectrum $F_{\rm\nu}$ is defined as the temperature of a blackbody whose spectrum has the same mean frequency ($\bar\nu$) as the observed spectrum \citep{MyersLadd1993}, and can be derived using the following expression: 
\begin{equation}
T_{\rm bol} = 1.25\times\,10^{-11} \bar\nu \,(K Hz^{-1})
\end{equation}
where $\bar\nu$ is defined as the ratio of the first and the zeroth frequency moments of the spectrum \citep{Ladd1991}:

\begin{equation}
\bar\nu = \frac{\int_{0}^{\nu_{max}} \,\nu \, F_\nu \,d\nu}{\int_{0}^{\nu_{max}} \, F_\nu \,d\nu}
\end{equation}

Basically, $\bar\nu$ provides a measurement of the {\em redness} of a given SED. As in the case of \lbol, it is necessary that the object shows detections in three different photometric tables to compute the bolometric temperature.\\

Both, $L_{\rm bol}$ and $T_{\rm bol}$ are estimated with a simple numerical integration using  photometric points from the optical to the sub-millimeter range. 

{\bf Dust\_properties:} The second table of the DB with physical parameters contains the estimated envelope masses (dust and dust+gas masses) of the objects, as quoted in the literature. We have also included the bibliographic reference from which these masses have been adopted. 

The values of the envelope dust masses in the literature have been mainly estimated from a given (sub-)mm continuum flux density (S$_{\lambda}$) at a given wavelength assuming optically thin conditions, and  following the equation:
\begin{equation}
M_{\mathrm{dust}}\,(M_{\odot})= \frac{S_{\lambda} D^2}{\kappa_{\lambda}\,B_{\lambda}(T_{dust})}
\end{equation}

where $D$ is the distance to the source (in pc), $B_{\lambda}$ is the Planck function, $T_{\rm dust}$ is the dust temperature in Kelvin, and $\kappa_{\lambda}$ is the opacity per dust+gas mass (in cm$^2$/g) at the observing wavelength.  Note that this value is normally estimated by the interpolation of dust opacities from theoretical models \citep{Ossenkopf1994}.
For completeness,  we have included in the table the dust temperature and the opacity used for the mass estimation, as quoted in the corresponding publication. In addition, we have included the observing wavelength,  and a field called \texttt{Beam} which is the observational beam size, which is useful to understand if the mass estimation comes from a single dish (single value) or interferometric observations (synthesized beam size in XY).
Note that for the objects with revised distances, we have recalculated the published masses using the new distance values.

For a few objects, the masses have been estimated either modeling the SED (e.g. ObjID~3) or using a particular model for the envelope (e.g. ObjID 204). In these cases, we have added a note in the 'Publication' field with this information, and we have only included in the table the $T_{\rm dust}$ and the dust mass resulting from the fit, without providing a particular wavelength or beam size  for the observations (a value of {\bf None} is included in those two fields). 


The total (gas+dust) masses in SUCANES are simply estimated from the dust mass, and assuming a gas-to-dust mass ratio of 100, so that $M_{\rm total}$\,=\,100\,$\times$\,$M_{\rm dust}$.  They are expressed in $M_{\odot}$.

The contents of the two tables including the derived $L_{\rm int}, L_{\rm bol}, T_{\rm bol}$ and the adopted $M_{\rm dust}$ \& $M_{\rm total}$, are described in Appendix~\ref{ap_tab}.

\subsection{SUCANES Molecular line emission table}

While continuum observations of very young substellar objects provide information about their dust content, the observations in different gas molecules are relevant to further characterize the objects, revealing important properties like e.g. the presence of molecular outflows, their chemical composition, or their belonging to a particular star forming region by analyzing their systemic velocities. 

A large fraction of SUCANES objects have observations in molecular spectral lines. Since these observations are very inhomogeneous (e.g. different antennas/interferometers, molecular transitions), we have constructed a table providing information about the availability of gas observations but without providing fluxes. The table contains a list of gas molecules frequently observed when studying embedded objects, together with the bibliographic references where these molecules have been presented. For each molecule, we have included a value of 0 if no data is available, and 1 if there is data, regardless of the detection/non-detection of the object. In particular, we have included information about the availability of CO, C, HCN, HCO$^{+}$, N$_{2}$H$^{+}$, CS, CN, CH$_{3}$OH and SiO. The particular transition for each molecule should be checked in the publications provided in the \texttt{pub\_mol} column.

In addition to the studied molecules, we have also provided information about 
the presence of molecular outflows  (1 for detected, 0 for non-detected, and {\bf None} if no data is available) associated to the objects, and the molecule used for the study.  
The bibliographic reference(s) from which these data have been retrieved are included in the last column. A detailed description of this table can be found in Appendix~\ref{ap_tab}.

\begin{figure}[t]
 \includegraphics[width=0.5\textwidth]
 {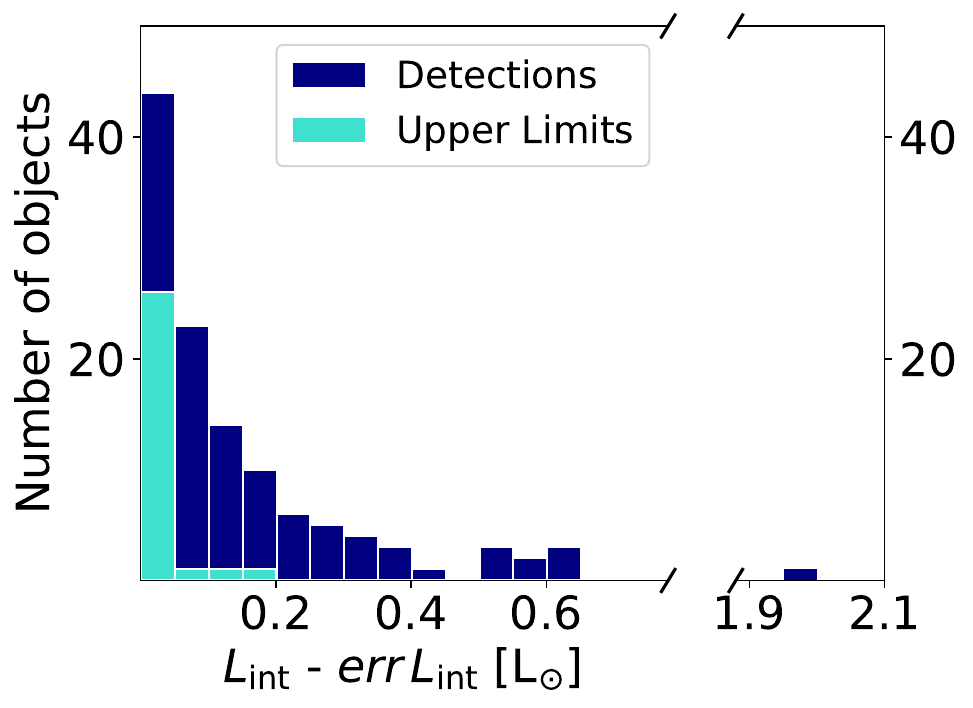}
\caption{Distribution of \lint-err\lint\, of the SUCANES objects. We have included detections and upper limits.}
\label{fig:Lint}
\end{figure}

\section{The SUCANES Graphical User Interface}\label{GraphInt}

The access to SUCANES by any external user should  be done through a Graphical User Interface (GUI) in the public webpage: 
\url{https://sucanes.cab.inta-csic.es/}. The SUCANES GUI allows any user to do queries, to download the requested information in \texttt{csv} files, and to plot the data in different graphics. In fact, any user can have access to all the data and physical parameters described in Section~\ref{db1} through this GUI,
allowing them to retrieve all the information contained in the eleven SUCANES tables in a single \texttt{csv} table.

In brief, the queries in the GUI can be done using the 'search' button at the top of the page. The searches can be done including either the name of an object (or part of it), the coordinates and/or an object type (e.g. proto-BD, pre-BD), that is, the user can select only by name, only by coordinates (RA and dec in degrees) or only by type,  or using both name and type, or both coordinates and type. The only combination that is not allowed is name and coordinates. The output of a query, if there are objects in SUCANES that fulfill the search inputs, is a html table showing basic information about the object(s): name, celestial coordinates, sky region, type and class. In addition, a boolean column (molecular emission) indicates if the object has been observed in any molecular transition. 
The user can also plot the SED of any of the objects in this table by clicking of the \texttt{sed} button located at the end of the row of each object. The SED is represented with the X-axis displaying the wavelength (in microns) and the Y-axis the flux density (in mJy), showing all the available data from the optical to the centimeter range. 

For any selected object, the user can download all the available data from SUCANES in a csv file, which will contain basic information about the object, the photometry to build the SED, the estimated envelope masses, and the derived \lint, \lbol, and \tbol\, parameters. An additional column will indicate if the object is included in the SUCANES molecular line emission table. The file can be retrieved by clicking on the \texttt{Photometry} button located at the the top right of the webpage. 

Finally, note that if the users are interested in downloading the whole database, they can do so by selecting 'ALL' in the \texttt{Types} field, click 'search', and then click on the \texttt{Photometry} button. As a result, the user will get a csv table containing the 174 objects with their main properties, the corresponding photometry, and all the derived physical parameters.

For a detailed description of SUCANES GUI, we refer to the "SUCANES User Manual": 
\url{https://sucanes.cab.inta-csic.es/documentation/}.

\section{Analysis of the SUCANES contents}\label{Contents}

  We have found a total of 174 objects in the literature classified as explained in Section~\ref{db1}. They are all included in Table~\ref{objects_SUCANES}. Fig. \ref{fig:hist1} summarizes their main properties, like e.g. their distances, the star forming region (or association/cloud) they belong to, their published classes and types.
 
The class of the SUCANES objects reflects a majority of Class~I sources that could be consistent with a proto-BD classification. 
A total of 31 objects do not have an assigned class: they correspond to the pre-BD (prestellar core analogs) and FHC candidates for which a class cannot be derived.
In the case of the types (Fig.\ref{fig:hist1}, bottom right panel), there is a majority of objects classified as VeLLOs.

Most of the SUCANES objects have been detected in nearby star-forming regions at distances smaller than $\sim$500\,pc, with a majority located at distances of $\sim$140\,pc and 436\,pc. In fact, there are two large groups of objects that belong to the Taurus and Aquila clouds, while the rest is more dispersed in different regions.  To illustrate this in detail, we have represented in Fig.~\ref{fig:hist2} the number of substellar candidates per star forming region, showing their \texttt{Class} (left panel) and their \texttt{Type} (right panel). As seen in the Figure, most of the sources in each group are Class I objects classified either as proto-BDs or VeLLOs. Taurus is the region with more objects of this type, while Chamaeleon\,II is the region with the largest number of identified pre-BD candidates.

In the next subsections, we will briefly discuss the physical properties of the SUCANES objects, i.e. \lint, \Menv, \lbol\,and \tbol. However, it should be noted that a deeper analysis of the SUCANES contents is presented in \cite{Palau2024_review}.

\subsection{Internal Luminosities}\label{sec:lint}

 Most of the substellar candidates in SUCANES have been identified using as a criterion the value of the $L_{\rm int}$ parameter proposed by \citet{Dunham08}. From the 174 objects in SUCANES, 149 have an estimation of the \lint: 120 show detections, and 29 upper limits. These 149 objects are represented in Fig.~\ref{fig:Lint}, where we show the values of \lint-err\lint used to identify VeLLOs.

\begin{figure}[t!]
 \includegraphics[width=0.5\textwidth]{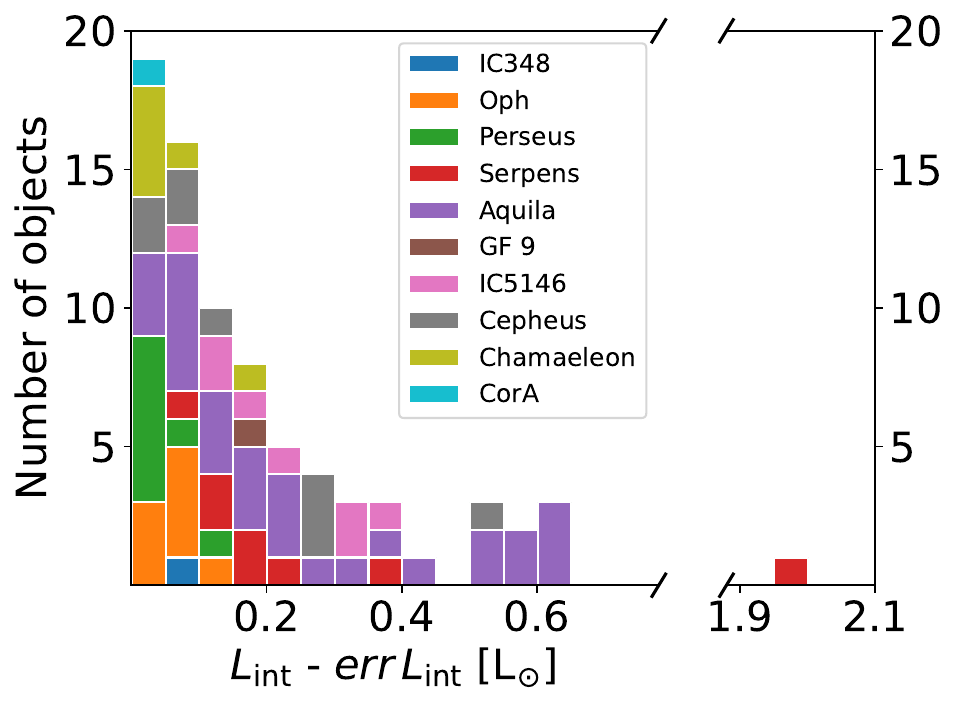}
\caption{Stacked histogram displaying the estimated \lint-err\lint\,for SUCANES objects with revised distances. As seen, several VeLLO candidates (mainly from Aquila) display \lint values above 0.2\,\lo\,within the errors.}
\label{fig:Lint_new}
\end{figure}

As seen in Fig.~\ref{fig:Lint}, several objects show values significantly larger than $\sim$ 0.2\,\lo\ within the errors.  This is mainly the result of using revised distances to some of the regions with more candidates, like e.g. Aquila or IC\,5146.
To illustrate this, we have represented in Fig. \ref{fig:Lint_new} the 
\lint\,-err\,\lint\,values estimated only for the objects with revised distances: in the case of the 28 VeLLOs in Aquila identified by \citet{Kim2016}, only 14 remain with \lint $\leq$ 0.2\,\lo. In the case of Serpens, SUCANES includes 14 objects, but only 8 have \lint\, estimations: three of them were studied by \citet{Riaz2016} assuming a distance of 260\,pc, and only one keeps now a \lint below 0.20\,\lo within the errors. One additional object (J183002) from \citet{Riaz2018} also shows a \lint\,$>$ 0.2\,\lo\,according to the Herschel flux at 70\,$\mu$m.  The rest of the objects remain as VeLLOs. Finally, for IC\,5146 and Cepheus, with eight and nine VeLLO candidates respectively, four objects in each region are discarded. The rest of objects in regions with revised distances remain as VeLLOs. Note that in the case of $\sigma$-Ori, the objects do not have data at 70\,$\mu$m and, therefore, they do not have \lint estimations. 


%

Since the revised distances allow us to conclude that several objects in SUCANES do not longer classify as VeLLOs or Proto-BD based on the \lint value, we have included a flag in the 'Type' column: these 25 objects are identified now with 'VeLLO/D' or 'Proto-BD/D', where 'D' refers to 'discarded'.

Apart from the 25 objects discarded due to their revised distances, there are three additional objects displaying  high \lint\, values not consistent with VeLLOs: two of them are in Taurus, IRAS\,04248+2612  and IRAS\,04191+1523\,B (ObjID 199, and 201, respectively), and one in Lupus-I, IRAS15398-3359 (ObjID 205).
The two proto-BD candidates in Taurus are part of  multiple systems:
IRAS\,04248+2612 is a triple system composed by a close binary (0\farcs16) and a wider companion at a separation of 4\farcs55, while IRAS\,04191+1523\,B is part of a binary system separated by 6\farcs09. 
For the two objects, the PACS70 $\mu$m photometry is therefore contaminated by the companions \citep[see][]{Bulger2014}, resulting in overestimated \lint\, values. Note that IRAS\,04191+1523\,B has been studied in detail with ALMA, showing an estimated dynamical mass of 0.12$\pm$0.01\,\mo \citep{Lee2017_iras04191B}, so we can also discarded as a proto-BD candidate.
In the case of IRAS\,15398-3359 in Lupus, with \lint=1.35\,\lo, it has an estimation of a dynamical mass of 7\,\mj\, \citep{Okoda2018}, so it is a bona-fide proto-BD candidate. The high value of \lint is interpreted as a result of an accretion burst \citep{Joergensen2013_I15398}. 
We have included a brief description of the three objects in Appendix \ref{ap_indiv_sources}.

The 25 objects (out of the 174) without \lint\,estimations are mainly pre-BD candidates (14 objects) that have been only detected in a single sub-millimeter band and have estimated (dust+gas) envelope masses below $\sim$\,80\mj\, 
\citep[e.g.][]{deGregorio2016, Huelamo17, Santamaria21}. The 11 objects left, either do not have observations available at 70 $\mu$m, or they have, but are undetected and there are no published upper limits.

To summarize, from the 174 objects included in SUCANES, we have finally kept a total of 121 objects with \lint $\leq$0.20\,\lo (both detections and upper limits). Two additional objects with higher values of \lint\,are still good substellar candidates given their properties. 
A total of 25 objects do not have \lint\,available data.


\begin{figure}[t]
\includegraphics[width=0.5\textwidth]
{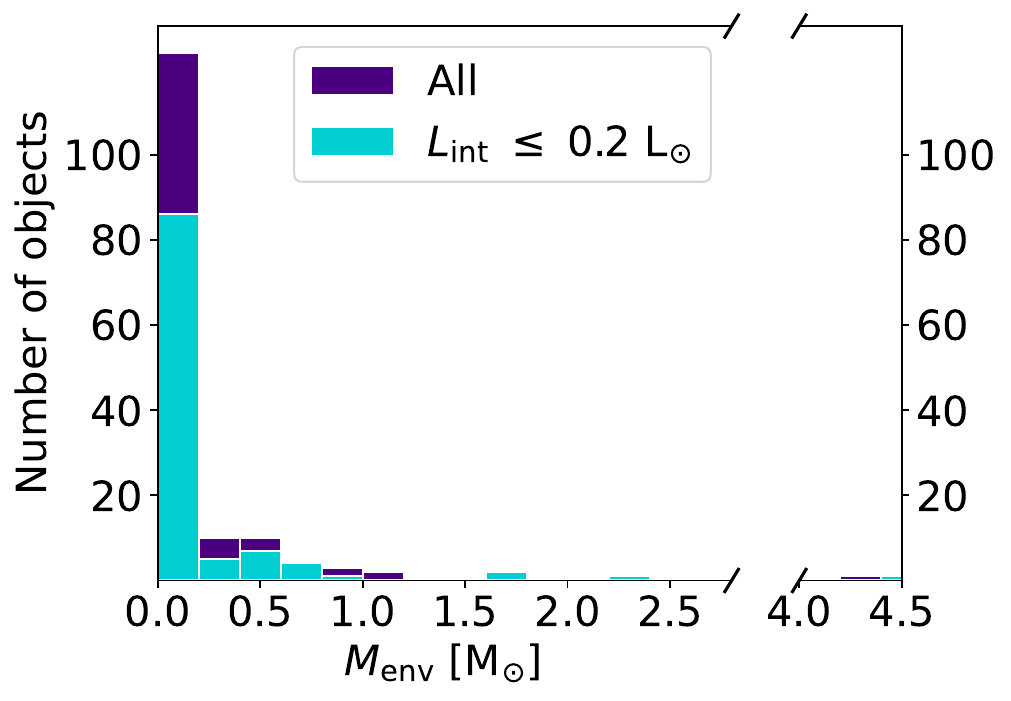}
\caption{Distribution of envelope masses for all the SUCANES objects. We have highlighted in cyan color the 107 VeLLOs with \lint\,$\leq$\,0.2\lo and envelope mass estimations.
}
\label{fig:menv}
\end{figure}

\subsection{Envelope Masses}

From the 174 objects included in SUCANES, we have retrieved envelope masses (total mass of gas+dust) for 158 of them. They are represented in  Fig.~\ref{fig:menv}.  
Most of the envelope masses (89 objects) have been derived from  Herschel/SPIRE flux at 250 $\mu$m \citep[objects from][]{Kim2016}. For the 69 left, 35 have masses estimated from 0.85-0.88\,mm data, 22 from 1.1-1.3\,mm data, and 4 from 350\,$\mu$m data. Note that only 6 objects have masses estimated from interferometric observations \citep[six pre-BDs from][] {Santamaria21}. For eight objects, the masses have been estimated through SED or envelope modeling. 

 From the 121 VeLLOs isolated in Section~\ref{sec:lint}, 107 have estimated envelope masses. They have been highlighted in Fig.~\ref{fig:menv}. As seen, a large fraction of them show \Menv\, $<$ 0.1\,\mo. This is an important parameter since it represents the mass reservoir from which the object will accrete material: depending on the total mass that is gathered, the object will end up as a brown dwarf or as a star.  The total mass accreted by the object can be estimated assuming a core formation efficiency (CFE) in low-mass dense cores. Different works have suggested very different values for this CFE, that can range between 10-50\% \citep[see e.g.][]{Motte1998, Alves2007, Bontemps2010, Palau2013}.  
Hence, although it is uncertain how much of the envelope mass will be finally accreted by the object, a low value of \Menv\, is required to guarantee that the proto-BD candidate does not reach a stellar mass. 

\begin{figure*}[t]
    \centering
\includegraphics[width=1\textwidth]
{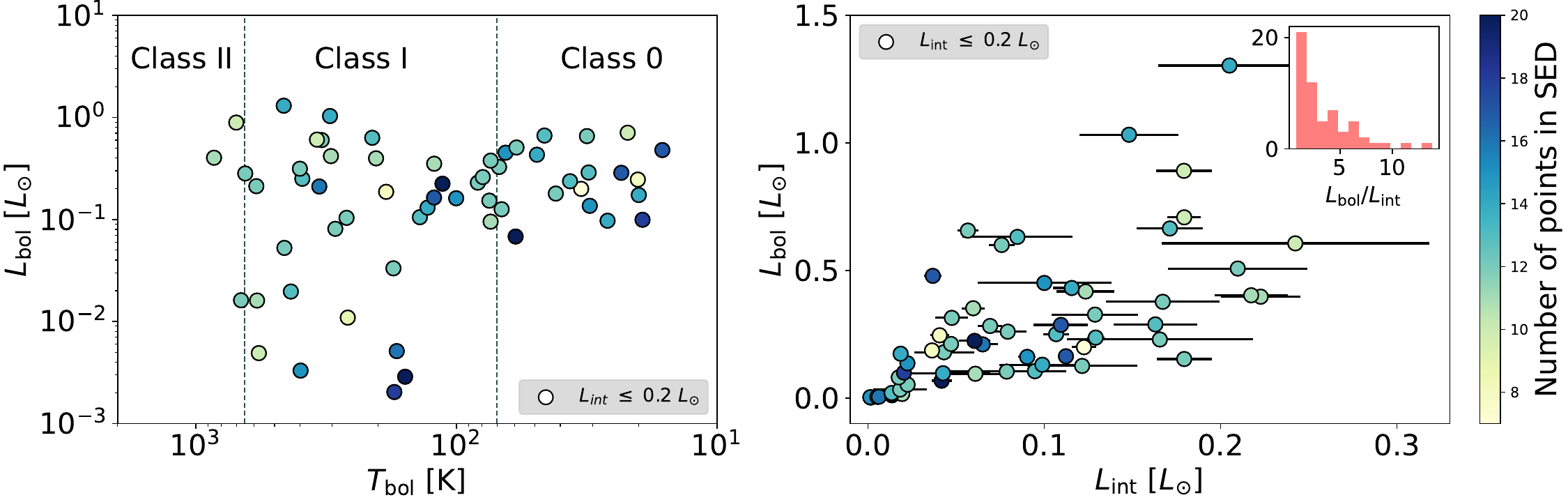}
\caption{{\bf Left:} \lbol-\tbol diagram with the 59 objects from SUCANES with well sampled SEDs and \lint $\leq$ 0.2 \lo (circles). The dotted vertical lines represent 
the \tbol\, values used to classify Class 0/I/II objects according to \citet{Chen1995}; 
{\bf Right:} \lbol\,versus \lint for the SUCANES objects with \lint\,$\leq$ 0.2 \lo. 
The inset histogram represents the distribution of \lbol/\lint ratios for the represented objects.
The colors of the circles in the two panels represent the number of data points used to build the SEDs of the objects.
}\label{fig:lboltbol}
\end{figure*}

 The four objects with \lint $<$ 0.2\,\lo\, and $M_{\rm env}$ $>$ 1\,\mo\, are Per-Bolo-58, Cha\,MMS-1 (see Appendix~\ref{ap_indiv_sources}),  J183014.4-013333, a VeLLO in Aquila identified by \citet{Kim2016},  and SM1-A, a proto-BD candidate in the Ophiuchus cloud (objID's 191, 192, 22, and 325, respectively). 
  \citep{Kawabe2018}. 
While Cha\,MMS-1 shows properties consistent both with a FHC or a Class~0 object \citep{Maureira2020_FHCs},  the other three are Class~0 objects. Hence, the four are in a very early stage of their formation; they still have to accrete a significant fraction of their masses, and  will probably end up as stars and not as BDs.

\subsection{\lbol - \tbol and \lbol - \lint diagrams}\label{sec:ltbol}

From the 174 objects in SUCANES, we have obtained \lbol and \tbol values for 85 objects with well-sampled SEDs. As explained in Section \ref{subsec:phys}, this  means that the objects have data in at least three photometric tables. In practice, we see that most of these objects show data between 3.6 to 850\,$\mu$m, with an average of $\sim$13 datapoints to build their SEDs. From these 85 objects, 59 show \lint\,$\le$\,0.2\,\lo. We have represented this subsample in Fig. \ref{fig:lboltbol} (left panel), indicating with colors the number of datapoints included in their SEDs.  As expected, most of the objects in SUCANES show \tbol\, values below $\sim$650\,K which corresponds to an evolutionary stage earlier or equal to Class~I, according to the classification by \citet{Chen1995}.
There are only three showing slightly higher temperatures:
J110955 in Cha I with 700\,K, 
J190418 (in Corona Australis) with 671\,K, 
and J040134.3+411143 in California with 853\,K. 


The \lbol show values between 2$\times$10$^{-3}$--1.3\,\lo. We have compared these \lbol\, with the estimated \lint\, in  Fig.~\ref{fig:lboltbol} (right panel). As expected, \lbol\,is higher than \lint: this is logical since in embedded  objects \lbol\,  is the sum of both the internal and the external luminosity (\lbol\, = \lint\, + \,$L_{\rm ext}$), where the external luminosity arises from the heating of the circumstellar envelope by the interstellar radiation field. Hence, for a given \lint, the large scatter in \lbol values seen in the figure might reflect a different irradiation of the substellar cores.  Fig.~\ref{fig:lboltbol} also shows that it is not simple to estimate the \lint of the objects assuming an approximate percentage of the \lbol, given the high scatter between these two luminosities. 
To illustrate this, we have included inside the figure a histogram displaying the ratio of the two luminosities (\lbol/\lint) for the displayed sample. 
As seen, around half of the objects show ratios equal or below 2.5, with the other half displaying ratios above this value (the objects with the highest ratios should be investigated further to undesrtand the origin of these extreme ratios).

\section{Scientific application: future observations of SUCANES objects with ngVLA and SKA}\label{sec:SciApp}

One of the goals of SUCANES is to use the compiled data of pre- and proto-BD candidates to identify the best targets for their future characterization with forthcoming astronomical facilities. In that respect, two of the most powerful observatories that will be key to characterize large samples of proto-BD candidates are the next generation Very Large Array\footnote{\url{https://ngvla.nrao.edu/}} (ngVLA) and the Square Kilometer Array\footnote{\url{https://www.skao.int/}} (SKA), which will allow us to study radiojets at the earliest phases of substellar evolution.

If formed as low-mass stars, proto-BDs are expected to show thermal radiojets as those reported in protostars \citep[e.g.][]{Anglada1995, Furuya2003, Anglada2018}. They are also expected to follow well-known correlations reported in young stellar objects that connect stellar and radiojet properties \citep[e.g. the relation between \lbol and the luminosity at 3.6\,cm; e.g. ][]{Shirley2007,Anglada2018}. 

\begin{figure*}[t!]
    \centering
    \includegraphics[width=1\textwidth]{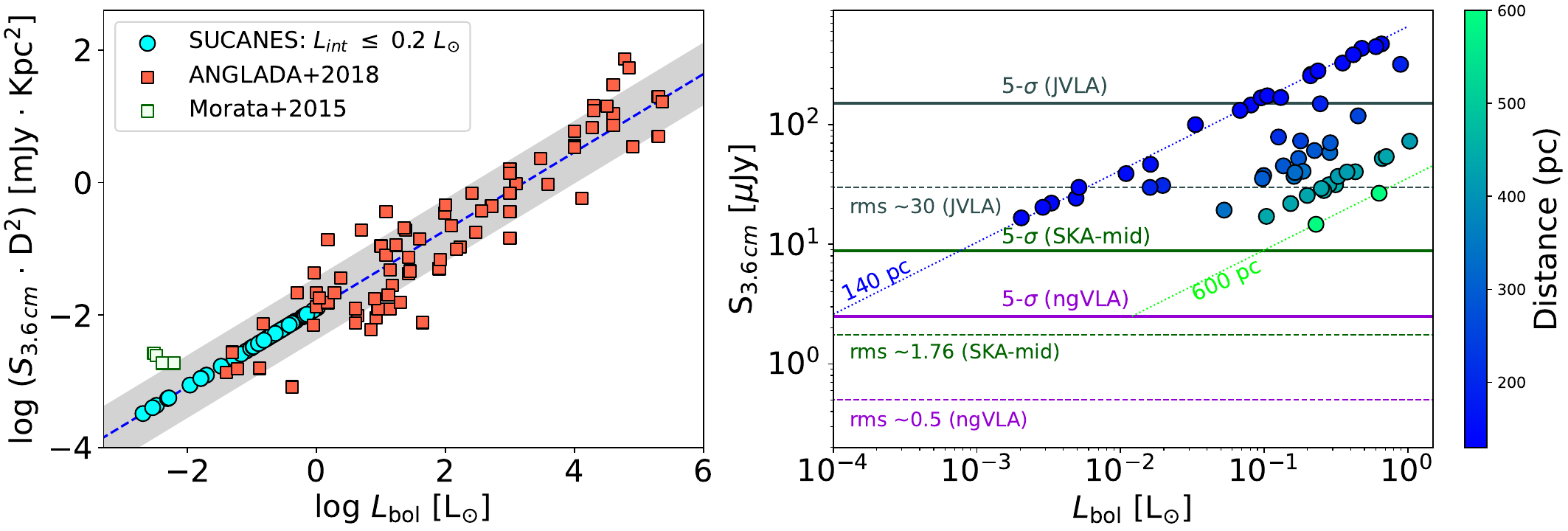}
\caption{{\bf Left:} 3.6\,cm luminosities of the substellar candidates with $L_{\rm int}$ $<$ 0.2\,\lo, and with an estimated $L_{\rm bol}$ in SUCANES (cyan circles). The 3.6\,cm luminosities have been derived using the correlation (blue dashed line) derived by \citet{Anglada2018} for YSOs (orange squares). We have marked with green squares the proto-BD candidates studied by \citet{Morata15}; {\bf Right:} Estimated 3.6\,cm flux densities of the SUCANES objects included in the left panel, using the distances included in our database. The grey lines represent the $rms$ in $\mu$Jy/beam (dashed) and the 5-$\sigma$ estimation (solid) from the JVLA observations included in \citet{Morata15}. The magenta and green lines are the expected sensitivities of the ngVLA and SKA-mid facilities, respectively, after 6 minutes of exposure time on-source: the dashed and solid lines represent the $rms$ (in $\mu$Jy/beam) and 5-$\sigma$ estimations, respectively. The datapoints are color-coded depending on their distances (see colorbar). The dotted lines represent objects at 140 pc (blue)  and 600\,pc (light green), revealing that ngVLA will be sensitive to very faint proto-BD candidates with \lbol$\sim$10$^{-4}$\lo and 10$^{-2}$\lo\,at these distances, respectively. 
}\label{fig:centimeter}
\end{figure*}

Radiojets are studied through centimeter observations: they can be either spatially resolved in a single band, or inferred using the spectral index from two centimeter bands. The first detections of thermal radiojets in a sample of proto-BD candidates were reported by \citet{Morata15} through Jansky Very Large Array (JVLA) observations at 1.3\,cm and 3.6\,cm. They observed a sample of 11 proto-BD candidates in the Barnard\,213 cloud in Taurus, and derived spectral indexes consistent with thermal radiojets in four sources. 
The detection of these four proto-BD candidates allowed \citet{Morata15} to extend the \lbol-$L_{3.6\,cm}$ relation to very low bolometric luminosities. Using the latest update of that relationship, \citep{Anglada2018} showed that the proto-BD candidates seem to follow the relation found in more massive YSOs
as displayed in Fig. \ref{fig:centimeter} (left panel). 
However, the figure also reflects the need to increase the number of proto-BD observations at centimeter wavelengths to understand if radiojets are common in these objects, and if so, to extract their main properties. 

The intrinsic faintness of proto-BD candidates limits the studies of radiojets. This will change in the near future with the advent of 
the ngVLA and the SKA observatories. These two facilities will provide superb sensitivity and angular resolution at centimeter wavelengths and, therefore, will be unique facilities to detect and characterize large samples of proto-BD candidates. 

As a preparation for future centimeter observations of proto-BD candidates, we have selected those objects in SUCANES with \lint $\le$ 0.2\,\lo and with available $L_{\rm bol}$ values, and estimate the expected 3.6\,cm luminosities assuming they follow
the correlation 
derived by \citet{Anglada2018} for young stellar objects:

\begin{equation}\label{anglada}
\mathrm{\left(\frac{S_{\nu}\cdot D^2}{mJy\,kpc^2}\right)} = 10^{-1.90\pm0.07} \,\left( \frac{L_{\rm bol}}{L_{\odot}}\right)^{0.59\pm0.03}
\end{equation}

The result is shown in Figure \ref{fig:centimeter} (left panel), where the SUCANES objects occupy the lower left corner of the diagram. Using the derived luminosity at 3.6\,cm and the distance to the sources included in SUCANES, we have estimated the flux densities at 3.6\,cm. We have represented them as a function of the bolometric luminosity in  Figure~\ref{fig:centimeter} (right panel).
As seen, most of the VeLLOs show estimated 3.6\,cm flux densities fainter than 200\,$\mu$Jy, with the faintest object displaying a flux of $\sim$16 $\mu$Jy (Figure~\ref{fig:centimeter}, right).

The JVLA 3.6\,cm maps presented by \citet{Morata15} in Barnard\,213 reached an average $rms$ of  ~30 $\mu$Jy/beam in 4.5-9 minutes of observing time on-source. This sensitivity, and the corresponding 5-$\sigma$ estimation, are represented in Fig.\ref{fig:centimeter} with dotted and solid grey lines, respectively. The average synthesized beam was 2.2$\times$1\farcs8 for uniform weighting and ~3\farcs1 $\times$ 2\farcs5 for natural weighting. 

We have estimated the expected sensitivities of ngVLA and SKA at 3.6\,cm using the corresponding exposure time calculators (ETC, see below). In both cases, we have assumed a target elevation of 45 degrees, a PWV of 6\,mm and, for a direct comparison with the JLVA performance,  an exposure time (on-source) of 6 minutes: in the case of ngVLA\footnote{\url{https://gitlab.nrao.edu/vrosero/ngvla-sensitivity-calculator}}, we have selected the main array (214 antennas of 18\,m) in the BAND\,2 (3.5 - 12.3\,GHz) considering the full bandwidth of 
$\sim$8.8\,GHz, and native resolution (natural, no taper). We estimate an $rms$ of 0.5\,$\mu$Jy/beam in 6 minutes of observing time with 25\,mas resolution in the continuum. In the case of SKA, we have used the SKA-mid sensitivity calculator\footnote{\url{https://sensitivity-calculator.skao.int/mid}} with the AA4 configuration (133 antennas of 15\,m), and the Band\,5a (4.6 - 8.5\,GHz)  using the full bandwidth of 3.8\,GHz. We have selected natural image weighting and no tapering. The $rms$ obtained after 6-min of exposure time is 1.76\,$\mu$Jy/beam. 

The ngVLA and SKA estimations are included in Fig.~\ref{fig:centimeter}, where the purple and green lines represent the sensitivity (rms and 5-$\sigma$) for the two observatories, respectively.
As seen, with a relatively modest exposure time investment, ngVLA and SKA-mid will allow us to observe 100\,\% of the sample displayed in Figure~\ref{fig:centimeter} with a signal-to-noise ratio (SNR) $>>$ 5 (magenta and green solid lines). Since each observatory will be located in a different terrestrial hemisphere, they will allow to detect the full sample of young substellar candidates with a high SNR. 
Moreover, the extrapolation of the datapoints representing the objects located at similar distances to the left of the plot (i.e. low \lbol values)  indicates that SKA and ngVLA will be able to detect radiojets of extremely low luminous proto-BDs with 
\lbol$\sim$10$^{-4}$\lo at 140\,pc, 
and \lbol$\sim$10$^{-2}$\lo at 600\,pc.
In any case, and as shown in Section~\ref{sec:ltbol}, these \lbol\, estimations can imply \lint\,values 2-10 times fainter.

As explained above, the presence of radiojets can be inferred from the spectral index derived using two centimeter bands \citep[see e.g.][]{Anglada2018}. In the case of \citet{Morata15}, they performed observations at 1.3 and 3.6\,cm. The expected sensitivity of ngVLA at 1.3\,cm (23\,GHz), covered by the BAND\,4 (20.5 - 34\,GHz), after 5 minutes of exposure time is 0.6\,$\mu$Jy/beam. This is almost 27 times more sensitive than the observations obtained with the JVLA by \citet{Morata15}, with a similar observing time (16\,$\mu$Jy/beam for 5 minutes on-source). 
Taking into account that the 3 proto-BD candidates in Taurus detected at 1.3\,cm by the JVLA showed fluxes of 70-80 $\mu$Jy, ngVLA will definitely help to increase the number of substellar candidates detected at this particular wavelength with high SNR. The combination of 1.3\,cm and 3.6\,cm high sensitive observations will provide spectral indexes for large samples of proto-BD candidates. Finally, note that although SKA will not initially observe at 1.3\,cm, the team will consider a future upgrade covering frequencies up to 24\,GHz.

As it follows, future studies of proto-BD candidates at centimeter wavelengths with ngVLA and SKA will allow us to confirm the true nature of a large sample of very young substellar candidates, and to characterize their mass-loss processes.




\section{Conclusions}\label{conclusions}

We have built the SUCANES database,  containing a list of objects classified as VeLLOs or substellar candidates at their earliest stages of formation in the literature. The database is a compilation of bibliographic data, and has been complemented with photometric data from public archives and catalogues. 

We have compiled a total of 174 objects, and build graphical tools to represent some of their properties.  We have analyzed the contents of the database, and reevaluated the classification of some objects after updating their distances and internal luminosities. 

SUCANES is designed as a compilation of sources with a bibliographic classification consistent with potential very young substellar candidates. Hence, it could be considered as a starting point to build subsamples of bona-fide substellar candidates \citep[as in][]{Palau2024_review}. We encourage the users to read the provided bibliographic references to understand the properties and classification of any particular source.

The database is public and open to the astrophysical community. Future upgrades of the DB include its implementation with new discoveries, 
and the revision of the nature of the objects based on results from new observations. If possible, we will work on the development of an automatic tool to allow any user to include newly discovered objects.

\section*{Acknowledgements}
\addcontentsline{toc}{section}{Acknowledgements}

This project has been funded by the European Space Agency\,-\,Science Faculty under contract No. 4000129603/19/ES/CM. We acknowledge CAB(CSIC-INTA) and ISDEFE for their support.
We are indebted to A. Parras and S. Suarez for their technical support, and to S. Cabañero and C. Valcarcel for the design of the SUCANES webpage and logo of the GUI. 
NH, DB, MMC, and MMH have been funded by the Spanish grants 
MCIN/AEI/10.13039/501100011033 PID2019-107061GB-C61 and PID2023-150468NB-I00.  
A.P. acknowledges financial support from the UNAM-PAPIIT IN113119 and IG100223 grants, the Sistema Nacional de Investigadores of CONAHCyT, and from the CONAHCyT project number 86372 of the `Ciencia de Frontera 2019’ program, entitled `Citlalc\'oatl: A multiscale study at the new frontier of the formation and early evolution of stars and planetary systems’, M\'exico. 
IdG acknowledges support from grant PID2020-114461GB-I00,funded by MCIN/AEI/10.13039/501100011033. 
OM is partially supported by the program Unidad de Excelencia Mar\'{\i}a de Maeztu, awarded to the Institut de Ci\`encies de l'Espai (CEX2020-001058-M). OM is supported by the European Research Council (ERC) under the European Union’s Horizon 2020 research and innovation programme (ERC Starting Grant "IMAGINE" No. 948582, PI: Daniele Vigan\`o).
AB acknowledges support from the Deutsche Forschungsgemeinschaft (DFG, German Research Foundation) under Germany's Excellence Strategy – EXC 2094 – 390783311. 
K.M. is funded by the European Union (ERC, WANDA, 101039452). Views and opinions expressed are however those of the author(s) only and do not necessarily reflect those of the European Union or the European Research Council Executive Agency. Neither the European Union nor the granting authority can be held responsible for them.
Any work using the database are encouraged to include this sentence in their acknowledgments: "This work has made use of the SUCANES database, a joint project funded by ESA Faculty under contract No. 4000129603/19/ES/CM, ISDEFE and CSIC."
\bibliographystyle{aa}
\bibliography{biblio} 

\begin{appendix}
\section{SUCANES DB tables}\label{ap_tab}

In this Appendix, we describe the contents of the tables that have been built for SUCANES. Note that the 'ObjID' is the unique identifier for each object in SUCANES Data Base, and all the tables are linked by this number. We describe the columns included in each table below:

\textbf{Main data tables:}
      \begin{itemize}
            \item Identity table:
      \begin{itemize}
            \item Identification number (ObjID)
            \item Name
            \item Sesame name 
       \end{itemize}
            \item Position table:
       \begin{itemize}
            \item Identification number (ObjID)
            \item RA and RA\_hms coordinates (degrees and sexagesimal)
            \item Decl and Decl\_dms (degrees and sexagesimal)
            \item Distance (pc)  
            \item eDistance: error in distance (pc)
            \item Type (VeLLO, pre-BD, proto-BD, Vello/FHC?, VeLLO/D, proto-BD/D, VeLLO/protostar) 
            \item Classification (0, I, 0/I, Flat) 
            \item Reference paper(s) 
      \end{itemize}
      \end{itemize}

\noindent\textbf{Photometric tables:}
      \begin{itemize}
            \item Optical and NIR photometry:
            \begin{itemize}
                  \item Identification number (ObjID)
                  \item Flux (Vega magnitude)
                  \item Flux error (Vega magnitude)
                  \item Filter Identification Number
                  \item Reference paper
            \end{itemize}    
            \item Spitzer and Herschel photometry:
            \begin{itemize}
                  \item Identification number (ObjID)
                  \item Flux (mJy)
                  \item Flux error (mJy)
                  \item Filter Identification Number
                  \item Reference paper
            \end{itemize}    
            \item Sub-mm and cm photometry:
            \begin{itemize}
                  \item Identification number (ObjID)
                  \item Variabilility (only for centimeter table)
                  \item Wavelength (micron)
                  \item Flux density (mJy)
                  \item Flux density error (mJy)
                  \item Peak Intensity (mJy/beam)
                  \item Peak intensity error (mJy/beam)
                  \item Integrated flux (mJy)
                  \item Integrated flux error (mJy)
                  \item Beam of observations (arcsec)
                  \item Position Angle of observations (degrees)
                  \item Angular Size of object in x direction (arcsec)
                  \item Angular Size of object in y direction (arcsec)
                  \item Resolved (0/1)
                  \item Position Angle (degrees)
                  \item Name of the Instrument
                  \item Reference paper
            \end{itemize}    

      \end{itemize} 
      
      An auxiliary table called {\bf Filters} describes the filters used for the photometric data: in each of the photometric tables described above (Optical, NIR, Spitzer, Herschel), we include a 'Filter Identification Number' after each flux value. This identification is defined as a  number that links each photometric filter with the auxiliary table that includes this information: 
            \begin{itemize}
            \renewcommand\labelitemi{--}
            \item Filter Identification Number
            \item Filter name
            \item Central wavelength (micron)
            \item FWHM (micron)
            \item Zero point Vega (Jy)
            \item Zero point Vega (erg/cm$^2$/s)
            \end{itemize}
\noindent\textbf{Derived physical parameters tables:}    

\begin{itemize}
    \item LumTbol table
             \begin{itemize}
                  \item Identification number (ObjID)
                  \item Internal luminosity (\lo)
                  \item Internal luminosity error (\lo)
                  \item Bolometric luminosity (\lo)
                  \item Bolometric temperature (K)
            \end{itemize}    

    \item Dust\_properties table
                \begin{itemize}
                  \item Identification number (ObjID)
                  \item Wavelength ($\mu$m)
                  \item Beam of observations (arcsec)
                  \item Dust temperature (K)
                  \item Opacity (cm$^2$g$^{-1}$)
                  \item Dust mass (\mo)
                  \item Total mass (\mo)
                  \item Reference paper
            \end{itemize}    
\end{itemize}

\noindent\textbf{Molecular lines table:}

\begin{itemize}
                  \item Identification number (ObjID)
                  \item CO available data: Yes (1)/ No (0)
                  \item C available data : Yes (1)/ No (0) 
                  \item HCN available data: Yes (1)/ No (0)
                  \item HCO$^{+}$ available data: Yes (1)/ No (0)
                  \item N$_2$H$^{+}$ available data: Yes (1)/ No (0)
                  \item CS available data: Yes (1)/ No (0) 
                  \item CN available data: Yes (1)/ No (0)
                  \item CH$_3$OH available data: Yes (1)/ No (0)
                  \item SiO available data: Yes (1)/ No (0)
                  \item Reference paper(s) 
                  \item Outflow detected: Yes (1)/ No (0)/ {\bf No information} ({\bf None})
                  \item Molecule used for outflow detection 
                  \item References for outflow detection 
\end{itemize}

For each molecule, we have included a value of 0 if no data is available, and 1 if there is data, regardless of the detection/non-detection of the object.
Note that, for simplicity, we have only included in the table the name of the studied molecule. To look for the particular transition that was observed, the user should check the bibliographic reference(s) provided for each source.

\section{Objects included in SUCANES}\label{ap:sucanes_objects}

 We summarize in Table~\ref{objects_SUCANES} all the objects included in SUCANES. We provide their main properties and the corresponding references.

%
\onecolumn

{\tiny
\begin{longtable}{lllllllll}
\caption{Objects included in SUCANES}\label{objects_SUCANES}\\
\hline\hline
Name & RA (J2000) & DEC (J2000) & Distance & Region & Type & Class & References & ObjID \\
 &  (hh:mm:ss.ss) & (dd:mm:ss.s) & (pc) &  &  & &  & \\
\hline
\endfirsthead
\caption{continued.}\\
\hline\hline
Name & RA (J2000) & DEC (J2000) & Distance & Region & Type & Class & Reference & ObjID \\
 &  (hh:mm:ss.ss) & (dd:mm:ss.s) & (pc) &  &  & &  &  \\
\hline
\endhead
\hline
\endfoot

%
  L1451-mm & 03:25:09.5 & +30:23:51.3 & 293$\pm$22 & Perseus & VeLLO/FHC? & N.A. & E06,P11,P14,M20 & 101\\
  L1448-IRS2E & 03:25:25.66 & +30:44:56.7 & 293$\pm$22 & Perseus & VeLLO & 0 & C10,T16,M20 & 195\\
  J032832.5+311105 & 03:28:32.5 & +31:11:05 & 293$\pm$22 & Perseus & Proto-BD & I & D08,K16,C16 & 0\\
  J032839.1+310601 & 03:28:39.1 & +31:06:01 & 293$\pm$22 & Perseus & Proto-BD & I & D08,K16,C16 & 1\\
  Per-Bolo45 & 03:29:07.7 & +31:17:16.8 & 293$\pm$22 & Perseus & VeLLO/FHC? & N.A. & E06,S12,T16,M20 & 176\\
  Per-Bolo58 & 03:29:25.7 & +31:28:16.3 & 293$\pm$22 & Perseus & VeLLO/Protostar & 0 & E10,M20 & 191\\
  J033032.6+302626 & 03:30:32.6 & +30:26:26 & 293$\pm$22 & Perseus & VeLLO/Protostar & I & D08,K16 & 2\\
  B1b-N & 03:33:21.2 & +31:07:43.8 & 293$\pm$22 & Perseus & VeLLO/FHC? & 0 & H14,G17,H19 & 173\\
  B1b-S & 03:33:21.4 & +31:07:26.4 & 293$\pm$22 & Perseus & VeLLO & 0 & H14,G17,H19 & 174\\
  IC348-SMMS2E & 03:43:57.73 & +32:03:10.1 & 321$\pm$10 & Perseus & Proto-BD & 0 & P14,K16,K19 & 3\\
  J040129.0+412128 & 04:01:29.0 & +41:21:28 & 450$\pm$23 & California & VeLLO & 0 & K16 & 44\\
  J040134.3+411143 & 04:01:34.3 & +41:11:43 & 450$\pm$23 & California & VeLLO & Flat & D15,K16 & 4\\
  IRAS04111+2800G & 04:14:12.3 & +28:08:37 & 137$\pm$10 & Taurus & Proto-BD & I & K16,K19 & 5\\
  J041726.38+273920.0 & 04:17:26.38 & +27:39:20.0 & 140$\pm$10 & Taurus & Proto-BD & I & P12,M15 & 166\\
  J041740.32+282415.5 & 04:17:40.32 & +28:24:15.5 & 140$\pm$10 & Taurus & Proto-BD & I & P12,M15 & 167\\
  J041757.7+274105 & 04:17:57.7 & +27:41:04.8 & 140$\pm$10 & Taurus & Proto-BD & I & B09,P12,M15 & 97\\
  J041757-NE & 04:18:00.30 & +27:41:36.3 & 137$\pm$10 & Taurus & Pre-BD & N.A. & P12 & 318\\
  J041828.08+274910.9 & 04:18:28.08 & +27:49:10.9 & 140$\pm$10 & Taurus & Proto-BD & I & P12,M15 & 168\\
  J041836.2+271442.2 & 04:18:36.28 & +27:14:42.6 & 140$\pm$10 & Taurus & Proto-BD & I & P12,M15 & 162\\
  J041840.2+282925 & 04:18:40.2 & +28:29:25 & 137$\pm$10 & Taurus & VeLLO & I & K16,K19 & 6\\
  J041847.84+274055.3 & 04:18:47.84 & +27:40:54.9 & 140$\pm$10 & Taurus & Proto-BD/D & I & P12,M15 & 163\\
  IRAS04158+2805 & 04:18:58.14 & +28:12:23.5 & 140$\pm$10 & Taurus & Proto-BD & I & W04,B14 & 198\\
  J041913.10+274726.0 & 04:19:13.10 & +27:47:26.0 & 140$\pm$10 & Taurus & Proto-BD & I & P12,M15 & 169\\
  J041938.7+282340.7 & 04:19:38.77 & +28:23:40.7 & 137$\pm$10 & Taurus & Proto-BD & I & P12,M15 & 164\\
  GKH94 41 & 04:19:46.57 & +27:12:55.21 & 140$\pm$10 & Taurus & Proto-BD & I & B14,DD16 & 202\\
  J042019.20+280610.3 & 04:20:19.20 & +28:06:10.3 & 140$\pm$10 & Taurus & Proto-BD & I & P12,M15 & 171\\
  J042118.43+280640.8 & 04:21:18.43 & +28:06:40.8 & 140$\pm$10 & Taurus & Proto-BD & I & P12,M15 & 172\\
  IRAM04191 & 04:21:56.90 & +15:29:46.4 & 140$\pm$10 & Taurus & Proto-BD & 0 & D08,K16,P14,K19 & 7\\
  IRAS04191+1523B & 04:22:00.44 & +15:30:21.21 & 140$\pm$10 & Taurus & Proto-BD/D & I & B14 & 201\\
  J042513.2+263145 & 04:25:13.2 & +26:31:45 & 137$\pm$10 & Taurus & VeLLO & 0 & K16,K19 & 45\\
  IRAS04248+2612 & 04:27:57.32 & +26:19:18.03 & 140$\pm$10 & Taurus & Proto-BD & I & W04,B14 & 199\\
  J042815.1+363028 & 04:28:15.1 & +36:30:28 & 450$\pm$23 & California & VeLLO & I & K16,K19 & 46\\
  J042818.6+365435 & 04:28:18.6 & +36:54:35 & 450$\pm$23 & California & VeLLO & I & K16 & 78\\
  L1521F-IRS & 04:28:38.90 & +26:51:35.6 & 137$\pm$10 & Taurus & Proto-BD & 0 & D08,K16,B14,P14,K19 & 8\\
  J043014.9+360008 & 04:30:14.9 & +36:00:08 & 450$\pm$23 & California & VeLLO & I & K16,K19 & 9\\
  J043055.9+345647 & 04:30:55.9 & +34:56:47 & 450$\pm$23 & California & VeLLO & I & K16,K19 & 47\\
  J043202.0+251641 & 04:32:02.0 & +25:16:41 & 137$\pm$10 & Taurus & VeLLO & I & K16 & 79\\
  J043411.5+240341 & 04:34:11.5 & +24:03:41 & 137$\pm$10 & Taurus & VeLLO & I & K16,K19 & 48\\
  J043909.0+261449 & 04:39:09.0 & +26:14:49 & 137$\pm$10 & Taurus & VeLLO & 0 & K16,K19 & 49\\
  J044022.4+255832 & 04:40:22.4 & +25:58:32 & 137$\pm$10 & Taurus & VeLLO & I & K16,K19 & 50\\
  IRAS04381+2540B & 04:41:12.69 & +25:46:36.08 & 140$\pm$10 & Taurus & Proto-BD & I & A05 & 204\\
  J044209.4+251635 & 04:42:09.4 & +25:16:35 & 137$\pm$10 & Taurus & VeLLO & I & K16,K19 & 51\\
  IRAS04489+3042 & 04:52:06.68 & +30:47:17.02 & 140$\pm$10 & Taurus & Proto-BD & I & W04,B14 & 200\\
  G192S & 05:29:54.35 & +12:16:29.68 & 380$\pm$42 & $\lambda$ Orionis & Proto-BD & 0/I & L16 & 203\\
  B30-LB10 & 05:31:09.29 & +12:11:08.8 & 400$\pm$40 & Barnard-30 & Pre-BD & N.A. & H17 & 209\\
  B30-LB31 & 05:31:15.32 & +12:03:38.2 & 400$\pm$40 & Barnard-30 & Pre-BD & N.A. & H17 & 210\\
  B30-LB08 & 05:31:22.97 & +12:11:34.7 & 400$\pm$40 & Barnard-30 & Pre-BD & N.A. & H17 & 208\\
  J053504.7-053712 & 05:35:04.7 & -05:37:12 & 420$\pm$42 & Orion & VeLLO & 0 & K16 & 52\\
  J053530.8-062632 & 05:35:30.8 & -06:26:32 & 420$\pm$42 & Orion & VeLLO & I & K16 & 53\\
  J053534.2-045952 & 05:35:34.2 & -04:59:52 & 420$\pm$42 & Orion & VeLLO & 0 & K16 & 80\\
  J053612.9-062330 & 05:36:12.9 & -06:23:30.6 & 420$\pm$42 & Orion & VeLLO & I & K16 & 54\\
  J053630.3-043216 & 05:36:30.3 & -04:32:16 & 420$\pm$42 & Orion & VeLLO & I & K16 & 55\\
  J053803.4-065815 & 05:38:03.4 & -06:58:15 & 420$\pm$42 & Orion & VeLLO & 0 & K16 & 56\\
  Mayrit1082188 & 05:38:34.45 & -02:53:51.46 & 402$\pm$25 & $\sigma$ Orionis & VeLLO & I & R15 & 207\\
  J054020.3-075114 & 05:40:20.3 & -07:51:14.9 & 420$\pm$42 & Orion & VeLLO & I & K16 & 81\\
  Mayrit1701117 & 05:40:25.79 & -02:48:55.42 & 402$\pm$25 & $\sigma$ Orionis & VeLLO & I & R15 & 206\\
  J054128.9-022319 & 05:41:28.9 & -02:23:19 & 420$\pm$42 & Orion & VeLLO & I & K16 & 82\\
  J054239.2-100147 & 05:42:39.2 & -10:01:47 & 420$\pm$42 & Orion & VeLLO & I & K16 & 83\\
  J055418.4+014903 & 05:54:18.4 & +01:49:03 & 420$\pm$42 & Orion & VeLLO & I & K16 & 57\\
  J080533.0-390924 & 08:05:33.0 & -39:09:24 & 440$\pm$100 & BHR16 & VeLLO & I & K16,K19 & 58\\
  J105959.7-771118 & 10:59:59.7 & -77:11:18 & 192$\pm$6 & Cha-I & VeLLO & Flat & K16,K19 & 84\\
  Cha-MMS1 & 11:06:31.7 & -77:23:33 & 192$\pm$6 & Cha-I & VeLLO/FHC? & N.A. & B06,T13,M20,B20 & 192\\
  J110955.0-763241 & 11:09:55.0 & -76:32:41 & 192$\pm$6 & Cha-I & VeLLO & Flat & K16,K19 & 59\\
  J121406.5-802625 & 12:14:06.5 & -80:26:25 & 193$\pm$12 & Cha-III & VeLLO & 0 & K16,K19 & 85\\
  ChaII-APEX-O & 12:51:39.0 & -77:08:55 & 198$\pm$6 & Cha-II & Pre-BD & N.A. & dG16 & 312\\
  ChaII-APEX-N & 12:51:55.8 & -77:08:43 & 198$\pm$6 & Cha-II & Pre-BD & N.A. & dG16 & 311\\
  ChaII-APEX-L & 12:52:55.8 & -77:07:35 & 198$\pm$6 & Cha-II & Proto-BD & 0 & dG16 & 165\\
  ChaII-APEX-A & 12:53:30.5 & -77:11:07 & 198$\pm$6 & Cha-II & Pre-BD & N.A. & dG16 & 302\\
  ChaII-APEX-E & 12:53:39.0 & -77:16:35 & 198$\pm$6 & Cha-II & Pre-BD & N.A. & dG16 & 305\\
  ChaII-APEX-J & 12:53:44.8 & -77:04:43 & 198$\pm$6 & Cha-II & Pre-BD & N.A. & dG16 & 310\\
  ChaII-APEX-G & 12:54:12.3 & -77:06:47 & 198$\pm$6 & Cha-II & Pre-BD & N.A. & dG16 & 307\\
  ChaII-APEX-B & 12:54:17.3 & -77:10:23 & 198$\pm$6 & Cha-II & Pre-BD & N.A. & dG16 & 303\\
  ChaII-APEX-F & 12:54:21.9 & -77:06:59 & 198$\pm$6 & Cha-II & Pre-BD & N.A. & dG16 & 306\\
  ChaII-APEX-I & 12:54:27.7 & -77:04:27 & 198$\pm$6 & Cha-II & Pre-BD & N.A. & dG16 & 309\\
  ChaII-APEX-H & 12:54:40.8 & -77:04:35 & 198$\pm$6 & Cha-II & Pre-BD & N.A. & dG16 & 308\\
  ChaII-APEX-C & 12:56:03.3 & -77:11:43 & 198$\pm$6 & Cha-II & Pre-BD & N.A. & dG16 & 304\\
  J125701.5-764834 & 12:57:01.5 & -76:48:34 & 198$\pm$6 & Cha-II & VeLLO & I & K16,K19 & 60\\
  ALMAJ153702.653-331924.92 & 15:37:02.65 & -33:19:24.92 & 94$\pm$4 & Lupus-I & Pre-BD & N.A. & SM21 & 177\\
  J153834.6-344819 & 15:38:34.6 & -34:48:19 & 150$\pm$20 & Lupus-I & VeLLO & I & K16 & 86\\
  ALMAJ153914.996-332907.62 & 15:39:14.99 & -33:29:07.62 & 153$\pm$5 & Lupus-I & Proto-BD & I & SM21 & 178\\
  J154051.6-342104 & 15:40:51.6 & -34:21:04 & 150$\pm$20 & Lupus-I & Proto-BD & I & D08,K16,K19 & 87\\
  L1YSO5 & 15:42:14.56 & -34:10:25.5 & 155$\pm$8 & Lupus-I & VeLLO & I & M17 & 327\\
  J154216.9-524802 & 15:42:16.9 & -52:48:02 & 250$\pm$50 & DC3272+18 & Proto-BD & 0 & D08,K16,K19 & 61\\
  ALMAJ154228.675-334230.18 & 15:42:28.67 & -33:42:30.18 & 153$\pm$5 & Lupus-I & Pre-BD & N.A. & SM21 & 179\\
  ALMAJ154229.778-334241.86 & 15:42:29.78 & -33:42:41.86 & 153$\pm$5 & Lupus-I & Proto-BD & 0/I & SM21 & 180\\
  IRAS15398-3359 & 15:43:02.21 & -34:09:07.71 & 150$\pm$20 & Lupus-I & Proto-BD & 0 & O14 & 205\\
  J154339.9-335554 & 15:43:39.9 & -33:55:54 & 150$\pm$20 & Lupus-I & VeLLO & I & K16 & 88\\
  ALMAJ154456.522-342532.99 & 15:44:56.52 & -34:25:32.99 & 153$\pm$5 & Lupus-I & Pre-BD & N.A. & SM21 & 181\\
  ALMAJ154458.061-342528.51 & 15:44:58.06 & -34:25:28.51 & 153$\pm$5 & Lupus-I & Pre-BD & N.A. & SM21 & 182\\
  ALMAJ154506.515-344326.15 & 15:45:06.52 & -34:43:26.15 & 153$\pm$5 & Lupus-I & Pre-BD & N.A. & SM21 & 183\\
  J154548.2-340510 & 15:45:48.2 & -34:05:10 & 150$\pm$20 & Lupus-I & VeLLO & I & K16 & 89\\
  ALMAJ154634.169-343301.90 & 15:46:34.17 & -34:33:01.90 & 153$\pm$5 & Lupus-I & Pre-BD & N.A. & SM21 & 184\\
  J160115.5-415235 & 16:01:15.5 & -41:52:35 & 150$\pm$20 & Lupus-IV & Proto-BD & I & D08,K16,K19 & 62\\
  ALMAJ160658.604-390407.88 & 16:06:58.60 & -39:04:07.88 & 155$\pm$10 & Lupus-III & Pre-BD & N.A. & SM21 & 185\\
  J160754.7-391544 & 16:07:54.7 & -39:15:44 & 200$\pm$20 & Lupus-III & VeLLO & I & K16 & 90\\
  ALMAJ160804.168-390452.84 & 16:08:04.17 & -39:04:52.84 & 155$\pm$10 & Lupus-III & Pre-BD & N.A. & SM21 & 186\\
  ALMAJ160920.089-384515.92 & 16:09:20.09 & -38:45:15.92 & 155$\pm$10 & Lupus-III & Pre-BD & N.A. & SM21 & 187\\
  ALMAJ160920.171-384456.40 & 16:09:20.17 & -38:44:56.40 & 155$\pm$10 & Lupus-III & Pre-BD & N.A. & SM21 & 188\\
  ALMAJ160932.167-390832.27 & 16:09:32.17 & -39:08:32.27 & 155$\pm$10 & Lupus-III & Pre-BD & N.A. & SM21 & 189\\
  ALMAJ161030.273-383154.52 & 16:10:30.27 & -38:31:54.52 & 155$\pm$10 & Lupus-III & Pre-BD & N.A. & SM21 & 190\\
  J162145.1-234231 & 16:21:45.1 & -23:42:31 & 138.4$\pm$2.6 & Ophiuchus & VeLLO & I & K16,K19 & 63\\
  Source-X & 16:26:27.43 & -24:24:18.26 & 137.0$\pm$1.2 & Ophiuchus & Proto-BD & 0 & K17,K18 & 326\\
  SM1-A & 16:26:27.86 & -24:23:59.56 & 137.0$\pm$1.2 & Ophiuchus & Proto-BD & 0 & K17,K18 & 325\\
  J162648.4-242838 & 16:26:48.4 & -24:28:38 & 138.4$\pm$2.6 & Ophiuchus & VeLLO & Flat & K16,K19 & 10\\
  OphB-11 & 16:27:13.96 & -24:28:29.3 & 138.4$\pm$2.6 & Ophiuchus & Pre-BD & N.A. & A12 & 197\\
  IRAS16253-2429 & 16:28:21.6 & -24:36:23 & 138.4$\pm$2.6 & Ophiuchus & Proto-BD & 0 & D08,K16,K19,H16 & 11\\
  MHO 2156 & 16:31:36.77 & -24:04:19.77 & 144.2$\pm$1.3 & Ophiuchus & Proto-BD & I & RB21 & 313\\
  ISO-OPH200 & 16:31:43.8 & -24:55:24.5 & 144.2$\pm$1.3 & Ophiuchus & Proto-BD & 0 & RM21 & 194\\
  SSTc2d J163152.5-245536 & 16:31:52.32 & -24:55:36.1 & 144.2$\pm$1.3 & Ophiuchus & Proto-BD & 0 & RB21 & 314\\
  J171941.24-265531 & 17:19:41.25 & -26:55:31.67 & 130$\pm$58 & Pipe Nebula & VeLLO & I & F09 & 193\\
  J180439.9-040122 & 18:04:39.9 & -04:01:22 & 436$\pm$9 & Aquila & VeLLO & I & K16,K19 & 64\\
  J180449.3-043639 & 18:04:49.3 & -04:36:39 & 436$\pm$9 & Aquila & VeLLO & I & K16 & 65\\
  J180941.9-033126 & 18:09:41.9 & -03:31:26 & 436$\pm$9 & Aquila & VeLLO & I & K16 & 66\\
  CB130-3-IRS & 18:16:15.6 & -02:32:45 & 436$\pm$9 & Aquila & Proto-BD & 0 & K16,K19 & 12\\
  L328-IRS & 18:16:59.47 & -18:02:30.5 & 217$\pm$30 & LDN328 & Proto-BD & 0 & L13,K16,K19,L18 & 13\\
  J182823.8-024952 & 18:28:23.8 & -02:49:52 & 436$\pm$9 & Aquila & VeLLO & 0 & K16 & 91\\
  SSTc2d J182844.8+005126 & 18:28:44.78 & +00:51:25.79 & 436$\pm$9 & Serpens & Proto-BD & 0/I & R18 & 320\\
  SSTc2d J182854.9+001813 & 18:28:54.90 & +00:18:32.68 & 436$\pm$9 & Serpens & Proto-BD & 0/I & R18 & 319\\
  J182855.78+002944.8 & 18:28:55.7 & +00:29:44.8 & 436$\pm$9 & Serpens & Proto-BD/D & 0 & R16 & 300\\
  J182855.8-013734 & 18:28:55.8 & -01:37:34 & 436$\pm$9 & Aquila & VeLLO/D & I & K16 & 14\\
  J182902.12+003120.7 & 18:29:02.12 & +00:31:20.7 & 436$\pm$9 & Serpens & Proto-BD & I & R16 & 196\\
  J182905.4-034245 & 18:29:05.4 & -03:42:45 & 436$\pm$9 & Aquila & VeLLO/D & I & K16,K19 & 15\\
  J182912.1-014845 & 18:29:12.1 & -01:48:45 & 436$\pm$9 & Aquila & VeLLO & I & K16,K19 & 67\\
  J182913.0-014617 & 18:29:13.0 & -01:46:17 & 436$\pm$9 & Aquila & VeLLO/D & I & K16,K19 & 16\\
  J182920.9-013714 & 18:29:20.9 & -01:37:14 & 436$\pm$9 & Aquila & Proto-BD/D & Flat & K16,K19 & 17\\
  J182925.1-014737 & 18:29:25.1 & -01:47:37 & 436$\pm$9 & Aquila & VeLLO & Flat & K16,K19 & 18\\
  SSTc2d J182927.4+003850 & 18:29:27.35 & +00:38:49.75 & 436$\pm$9 & Serpens & Proto-BD & 0/I & R18 & 323\\
  J182933.6-014510 & 18:29:33.6 & -01:45:10 & 436$\pm$9 & Aquila & VeLLO & Flat & K16,K19 & 19\\
  J182937.4-031453 & 18:29:37.4 & -03:14:53 & 436$\pm$9 & Aquila & VeLLO/D & I & K16,K19 & 68\\
  SSTc2d J182940.2+001513 & 18:29:40.20 & +00:15:13.11 & 436$\pm$9 & Serpens & Proto-BD & I & R18,R21 & 316\\
  J182943.9-021255 & 18:29:43.9 & -02:12:55 & 436$\pm$9 & Aquila & Proto-BD/D & I & K16,K19 & 20\\
  J182949.57+011706.0 & 18:29:49.5 & +01:17:06.0 & 436$\pm$9 & Serpens & Proto-BD/D & I & R16 & 301\\
  SSTc2d J182952.1+003644 & 18:29:52.06 & +00:36:43.63 & 436$\pm$9 & Serpens & Proto-BD & Flat & R18 & 324\\
  J182952.9-015805 & 18:29:52.9 & -01:58:05 & 436$\pm$9 & Aquila & VeLLO & 0 & K16,K19 & 21\\
  J182953.04+003606.8 & 18:29:53.04 & +00:36:06.8 & 436$\pm$9 & Serpens & Proto-BD & Flat & D08 & 142\\
  MHO 3256 & 18:29:57.66 & +01:13:04.6 & 436$\pm$9 & Serpens & Proto-BD & 0 & R21 & 315\\
  J182958.3-015740 & 18:29:58.3 & -01:57:40 & 436$\pm$9 & Aquila & VeLLO & I & K16,K19 & 69\\
  SSTc2d J182959.4+011041 & 18:29:59.38 & +01:10:41.08 & 436$\pm$9 & Serpens & Proto-BD & 0/I & R18 & 322\\
  SSTc2d J183002.1+011359 & 18:30:02.09 & +01:13:58.98 & 436$\pm$9 & Serpens & Proto-BD & 0/I & R18 & 321\\
  J183014.4-013333 & 18:30:14.4 & -01:33:33 & 436$\pm$9 & Aquila & VeLLO & 0 & K16,K19 & 22\\
  J183015.6-020719 & 18:30:15.6 & -02:07:19 & 436$\pm$9 & Aquila & VeLLO/D & 0 & K16,K19 & 23\\
  J183016.2-015252 & 18:30:16.2 & -01:52:52 & 436$\pm$9 & Aquila & VeLLO/D & 0 & K16,K19 & 25\\
  J183017.4-020958 & 18:30:17.4 & -02:09:58 & 436$\pm$9 & Aquila & VeLLO/D & I & K16,K19 & 24\\
  J183021.8-015201 & 18:30:21.8 & -01:52:01 & 436$\pm$9 & Aquila & VeLLO/D & I & K16,K19 & 26\\
  J183027.5-015439 & 18:30:27.5 & -01:54:39 & 436$\pm$9 & Aquila & VeLLO/D & I & K16,K19 & 27\\
  J183046.9-015645 & 18:30:46.9 & -01:56:45 & 436$\pm$9 & Aquila & VeLLO/D & 0/I & K16 & 28\\
  J183047.6-024356 & 18:30:47.6 & -02:43:56 & 436$\pm$9 & Aquila & VeLLO & 0 & K16 & 70\\
  J183237.4-025045 & 18:32:37.4 & -02:50:45 & 436$\pm$9 & Aquila & Proto-BD/D & I & K16,K19 & 29\\
  J183242.4-024756 & 18:32:42.4 & -02:47:56 & 436$\pm$9 & Aquila & VeLLO/D & Flat & K16,K19 & 30\\
  J183245.6-024657 & 18:32:45.6 & -02:46:57 & 436$\pm$9 & Aquila & VeLLO & I & K16,K19 & 31\\
  J183329.4-024558 & 18:33:29.4 & -02:45:58 & 436$\pm$9 & Aquila & VeLLO & 0 & K16,K19 & 32\\
  J183929.8+003740 & 18:39:29.8 & +00:37:40 & 436$\pm$9 & Serpens & VeLLO & I & K16,K19 & 33\\
  J183936.1-001151 & 18:39:36.1 & -00:11:51 & 436$\pm$9 & Serpens & VeLLO & I & K16 & 92\\
J190418.6-373556 & 19:04:18.6 & -37:35:56 & 154$\pm$4 & Corona Aus & VeLLO & I & K16 & 94\\
  L673-7-IRS & 19:21:34.82 & +11:21:23.4 & 300$\pm$100 & L673-7 & VeLLO/Protostar & 0 & K08,D10,K16,K19 & 34\\
  L1148-IRS & 20:40:56.66 & +67:23:04.9 & 330$\pm$1 & Cepheus & Proto-BD & I & D08,P14,K16,K19 & 35\\
  GF9-2 & 20:51:29.82 & +60:18:38.5 & 270$\pm$10 & GF9 & VeLLO & 0 & F06,P14,C18 & 96\\
  J210221.2+675420 & 21:02:21.2 & +67:54:20 & 341$\pm$2 & Cepheus & VeLLO/D & I & K16 & 317\\
  J210227.3+675418 & 21:02:27.3 & +67:54:18 & 341$\pm$2 & Cepheus & VeLLO & I & K16 & 37\\
  J210340.4+682631 & 21:03:40.4 & +68:26:31 & 341$\pm$2 & Cepheus & VeLLO & I & K16 & 93\\
  L1014-IRS & 21:24:07.5 & +49:59:09 & 258$\pm$50 & Cygnus & Proto-BD & 0 & H06,K08,D08,K16,K19 & 38\\
  J214448.3+474459 & 21:44:48.3 & +47:44:59.7 & 600$\pm$100 & IC5146 & VeLLO/D & 0 & K16,K19 & 71\\
  J214457.0+474152 & 21:44:57.0 & +47:41:52 & 600$\pm$100 & IC5146 & VeLLO & I & K16,K19 & 39\\
  J214531.2+473621 & 21:45:31.2 & +47:36:21 & 600$\pm$100 & IC5146 & VeLLO/D & I & K16,K19 & 40\\
  J214657.5+473223 & 21:46:57.5 & +47:32:23 & 600$\pm$100 & IC5146 & VeLLO & I & K16,K19 & 72\\
  J214703.0+473314 & 21:47:03.0 & +47:33:14 & 600$\pm$100 & IC5146 & VeLLO & I & K16,K19 & 73\\
  J214706.0+473939 & 21:47:06.0 & +47:39:39 & 600$\pm$100 & IC5146 & VeLLO & I & K16,K19 & 41\\
  J214755.6+473711 & 21:47:55.6 & +47:37:11 & 600$\pm$100 & IC5146 & VeLLO/D & I & K16,K19 & 74\\
  J214858.5+472542 & 21:48:58.5 & +47:25:42.8 & 600$\pm$100 & IC5146 & VeLLO/D & Flat & K16,K19 & 75\\
  J215607.3+764229 & 21:56:07.3 & +76:42:29 & 339$\pm$1 & Cepheus & VeLLO & Flat & K16,K19 & 76\\
  J222933.3+751316 & 22:29:33.3 & +75:13:16 & 339$\pm$1 & Cepheus & VeLLO & I & K16,K19 & 77\\
  J222959.4+751403 & 22:29:59.4 & +75:14:03 & 339$\pm$1 & Cepheus & VeLLO/D & I & K16,K19 & 42\\
  J223031.94+751408.9 & 22:30:31.94 & +75:14:08.9 & 339$\pm$1 & Cepheus & VeLLO/D & 0 & D08, L10, K08, W07 & 117\\
  L1251A-IRS4 & 22:31:05.5 & +75:13:37.2 & 339$\pm$1 & Cepheus & VeLLO/D & 0 & K16,K19 & 43\\ \hline
\end{longtable}
}
\tablebib{
  A99: \citet{Andre1999}; 
  A05: \citet{Apai2005_ProtoBD_NICMOS}; 
  A12: \citet{Andre12}; 
  B06: \citet{Belloche2006_ChaMMS1}; 
  B09: \citet{Barrado2009};   
  B14: \citet{Bulger2014}; 
  B20: \citet{Busch2020};
  C10: \citet{Chen2010_L1448_FHC};
  C16: \citet{Carney2016};
  C18: \citet{Clemens2018};
  D08: \citet{Dunham08}; 
  D15: \citet{Dunham2015_YSO_GouldBelt};
  DD16: \citet{Dang2016}; 
  E06: \citet{Enoch2006};
  E10: \citet{Enoch2010_FHC};  
  F06: \citet{Furuya2006_GF9-2}; 
  F09: \citet{Forbrich2009ApJ_Pipe}; 
  dG16: \citet{deGregorio2016}; 
  G17: \citet{Gerin2017}; 
  H06: \citet{Huard2006_L1014};
  H14: \citet{Hirano2014}; 
  H16: \citet{Hsieh2016_Binary_ProtoBD_Jet_IRAS16253};
  H17: \citet{Huelamo17}; 
  H19: \citet{Hirano2019_B1bNS}; 
  K08: \citet{Kauffmann2008};
  K16: \citet{Kim2016}; 
  K17: \citet{Kirk2017_Ophiuchus}; 
  K18: \citet{Kawabe2018};
  K19: \citet{Kim2019}; 
  L10: \citet{Lee2010_L1251};
  L13: \citet{LeeCW2013_L328}; 
  L16: \citet{Liu2016_Planck}; 
  L18: \citet{LeeCW2018}; 
  M15: \citet{Morata15}; 
  M17: \citet{Mowat2017};
  M20: \citet{Maureira2020_FHCs};
  O14: \citet{Oya2014}; 
  P11: \citet{Pineda2011_FHC_VeLLO_L1451};
  P12: \citet{Palau12}; 
  P14: \citet{Palau14};   
  R15: \citet{Riaz2015_VeLLos}; 
  R16: \citet{Riaz2016};
  R18: \citet{Riaz2018};
  RB21: \citet{RiazBally2021_AccretionOutflow_ProtoBD}; 
  RM21: \citet{RiazMachida2021_Structure_ProtoBD};  
  SM21: \citet{Santamaria21}; 
  T13: \citet{Tsitali2013}; 
  T16: \citet{Tobin2016_VANDAM}; 
  W04: \citet{White2004}; 
 W07: \citet{Wu2007_L1251}; 
}

\twocolumn

\section{Notes on individual sources included in SUCANES}\label{ap_indiv_sources}

In this Appendix, we briefly describe the properties of some of the sources included in SUCANES to explain their inclusion in the database: 

{\bf IRAM04191}: This Class~0 object was discovered by \citet{Andre1999}, and classified as a VeLLO by \citet{Dunham2006} after the analysis of {\em Spitzer} data. The object was included in the sample analyzed by \citet{Kim2016,Kim2019} who concluded that is probably a protostar based on its accretion properties, although this classification is highly uncertain, as explained by the authors. Note that there are indirect evidences of this object being a binary system \citep{Lee2005}. 



{\bf IRAS\,04191+1523\,B:} This object was classified as a proto-BD candidate by \citet{Bulger2014}. The target is the secondary component of a binary separated by 6$\farcs$1 \citep{Duchene2004}. 
\citet{Luhman2010_Taurus} estimated a spectral type of M6-M8, which implied that the central source is a substellar or a Very Low Mass object \citet{Dang2016}.
The photometry at wavelengths longer than 8$\mu$m is contaminated by the primary star emission, which can explain the SUCANES estimated values for $L_{\rm int}$ and $L_{\rm bol}$ (0.37 and 0.49\,$L_{\odot}$, respectively).
The dynamical mass of the object, derived through ALMA C$^{18}$O observations, is of 120\,\mj\, after correction from the disk inclination. Hence, the source properties are more consistent with a low-mass protostar and not with a proto-BD.

 {\bf {[GKH94]} 41}: \citet{Luhman2010_Taurus} classified the source as 'Class~I?', but the study of \citet{Furlan2011} concluded that this was a disk-dominated source and probably a more evolved object. However, the analysis of millimeter observations presented by \citet{Dang2016} revealed the presence of an envelope, confirming that this is a Class~I source. They estimate an upper limit to the mass of the central object in the range of 14 - 97\,$M_{\rm Jup}$.

{\bf IRAS\,04248+2612:} This object was classified as a proto-BD by \citet{White2004} based on its near-IR and optical properties. The authors reported a spectral type of M5.5 and a central mass of 0.07\,$M_{\odot}$, according to the temperature-mass relation from \citet{Siess2000} evolutionary models at 1\,Myr. 
The target is a triple system \citep{Duchene2007}, composed by a close binary (separation of 0\farcs16) and a third object
at a separation of  4\farcs55. \citet{Bulger2014} and \citet{Mottram2017} have presented Herschel 70 $\mu$m photometry of the unresolved system that results in a \lint of 0.32$\pm$0.04\lo. 
The estimated mass of the envelope is 0.25 $M_{\odot}$ \citep{Motte2001}.

{\bf IRAS\,04158+2805}: It was classified as a proto-BD by \citet{White2004}. However, a work by \citet{Luhman2006} suggested that the object is a low-mass star rather than a brown dwarf. Its evolutionary stage is controversial since it is surrounded by a high inclined disk that can block the stellar light thus mimicking a younger, embedded system \citep[see e.g. the discussion by][]{Ragusa2021}. 
However, other works show that its properties are more consistent with a Class~I source in Taurus \citep[e.g.][]{Furlan2008}. The object has been resolved into a 25\,au separation binary, and its total dynamical mass ranges between 0.15-0.45\,\mo. Depending on the mass of each component, it might host at least one substellar object.

{\bf IRAS\,04489+3042}: It was classified as a proto-BD by \citet{White2004}. However,  \citet{Luhman2006} suggested that the object is a low-mass star rather than a brown dwarf.

{\bf IRAS\,15398-3359:} This source was included in SUCANES since \citet{Oya2014} classified it as a substellar object. They  provided an upper limit to its mass of $<$0.09\,M$_{\odot}$ based on the modeling of its rotating infalling envelope.
\citet{Okoda2018} estimated a dynamical mass of 0.007 M$_{\odot}$ based on the fit to a Keplerian motion of the PV diagram of the SO molecule.  We estimate a \lint$\sim$1.3\,L$_{\odot}$, which is not consistent with a VeLLO. However, \citet{Joergensen2013_I15398} has suggested that such a large \lint could be the result of an accretion burst that took place during the last 100-1000 years.


  {\bf B1b-N \& B1b-S:} These two sources are located in the Barnard-1 cloud in Perseus \citep{Hirano1999}. \citet{Hirano2014}, derived \lint values consistent with VeLLOs. The new distance of 293$\pm$22\,pc assumed in SUCANES still provides \lint\, consistent with that classification ($<$0.01\lo \,and 0.05\,\lo for B1b-N and B1b-S, respectively). \citet{Hirano2014} concluded that the properties of the two sources are consistent with Class~0 objects less evolved than other Class~0 protostars, with B1-bN displaying properties consistent with a FHC. The study presented by \citet{Gerin2017} using ALMA reached similar conclusions. Hence, B1b-N has been classified as a VeLLO/FHC?, while B1b-S has classified it as 'VeLLO'. The estimated envelope mass is of 0.56\mo\,for both objects, using the updated distance of 293\,pc \citep{Hirano2014}. 






{\bf GF 9-2:}  Based on its upper limit to the \lint, \citet{Palau14} classified this object as a  VeLLO.  We have derived an upper limit to the \lint of $<$ 0.2\,\lo\,using the revised distance to GF\,9 of 270$\pm$10\,pc \citep{Clemens2018}. Note that \citet{Furuya2019} have discarded it as a proto-BD candidate based on the estimated envelope infall rate.

{\bf Cha MMS-1}: This object is embedded in a filament of the Chamaeleon~I cloud, and was classified as a possible FHC by \citet{Belloche2011}. The \lint derived in SUCANES is 0.02\lo. \citet{Tsitali2013} estimated the \lint\,based on a 3D Radiative magneto-hydrodynamical model to be between 0.13-0.29\,\lo\,(for an updated distance of 192\,pc). \citet{Busch2020} confirmed the presence of a bipolar outflow in the source through ALMA observations, and argued that its evolutionary stage is more consistent with a Class~0 object than with a FHC. \citet{Maureira2020_FHCs} presented ALMA observations concluding that although their data could not rule out the object as a FHC, the properties of its CO outflow and SED  were in better agreement with the predictions for a young protostar.
The mass of the envelope derived from 870\,$\mu$m single dish data is 5.5\,\mo, considering the total flux associated to the spatially resolved source \citep{Belloche2011}. Given this mass estimation, the object will probably end up as a star. We have included VeLLO/FHC? in its \texttt{Type} field.

{\bf L1448-IRS2E}: This object was first reported as a VeLLO by \citet{Chen2010_L1448_FHC}.  
An ALMA study by \citet{Maureira2020_FHCs} showed neither emission of dense gas (traced by N$_2$H$^{+}$, and NH$_2$D) at the position of the VeLLO nor detection of any source at the 3.3\,mm continuum. They concluded that there is not any core at the position of the target. Instead, they explained the emission reported by \citet{Chen2010_L1448_FHC} as gas heated by the outflow of the neighboring Class~0 source, L1448 IRS2. Hence, the nature of this object is doubtful.

VeLLOs included in Table~5 from \citet{Kim2019}: the 19 targets included in this table are VeLLOs that show outflows.
They have been classified either as proto-BD or protostars based on the estimation of the accreted mass and the mass of the envelope. However, this classification should be taken with caution given the associated uncertainties. As explained by the authors, "{\em the estimation of the accreted mass and
the envelope mass can be affected by various uncertain
parameters, which are mostly unknown, and hence our
classification of the sources can be highly uncertain}".
One of the objects classified as a possible protostar is IRAM04191 (see above) that we have classified as a proto-BD candidate based on its binarity properties.


{\bf J041847.84+274055.3}: The infrared object in Taurus (ObjID 163) has been recently identified as a galaxy by Bouy et al. (in prep.) using Euclid data in the visible band. Hence, we have discarded it as a proto-BD candidate, and included a 'D' in the type column.

\end{appendix}

\end{document}